\documentclass[12pt,preprint,psfig]{aastex}
\begin{document}
\title{Unified 1-D Simulations of Gamma-ray Line Emission from Type Ia Supernovae}
\author{P.A. Milne\footnote{Current address: Steward Observatory, Tucson, AZ 85721, 
Email: pmilne@as.arizona.edu}, A.L. Hungerford\footnote{Department of Astronomy, 
University of Arizona, Tucson, AZ 85721}, C.L. Fryer, T. M. Evans, T. J. Urbatsch}
\affil{Los Alamos National Laboratory, Los Alamos, NM 87545}
\author{S.E. Boggs}
\affil{Department of Physics, University of California, Berkeley, CA 94720}
%\author{P. H\H{o}flich}
%\affil{Department of Astronomy, University of Texas, Austin, TX 78712}
\author{J. Isern, E. Bravo, A. Hirschmann}
\affil{Institut d'Estudis Espacials de Catalunya (IEEC), Barcelona, Spain 08034}
\author{S. Kumagai}
\affil{College of Science and Technology, Nihon University, Tokyo, Japan 101-8308}
\author{P.A. Pinto}
\affil{Department of Astronomy, University of Arizona, Tucson, AZ 85721}
\author{L.-S. The}
\affil{Department of Physics and Astronomy, Clemson University, 
Clemson, SC 29634-1911} 


\begin{abstract}

The light curves of Type Ia Supernovae (SN Ia) are powered by
gamma-rays emitted by the decay of radioactive elements such as
$^{56}$Ni and its decay products.  These gamma-rays are downscattered,
absorbed, and eventually reprocessed into the optical emission which
makes up the bulk of all supernova observations.  Detection of the
gamma-rays that escape the expanding star provide the only direct
means to study this power source for SN Ia light curves.
Unfortunately, disagreements between calculations for the gamma-ray
lines have made it difficult to interpret any gamma-ray observations.
 
Here we present a detailed comparison of the major gamma-ray line
transport codes for a series of 1-dimensional Ia models.
Discrepancies in past results were due to errors in the codes,
and the corrected versions of the seven different codes yield
very similar results.  This convergence of the simulation results allows
us to infer more reliable information from the current set of
gamma-ray observations of SNe Ia.  The observations of SNe 1986G,
1991T and 1998bu are consistent with explosion models based on their
classification: sub-luminous, super-luminous and normally-luminous
respectively.

\end{abstract}

\section{Introduction}
                                                                                                            
Type Ia supernovae (SNe Ia) are intertwined with many of the most
interesting frontiers of astrophysics. They occur in all galaxy
types and are important contributors to
galactic chemical evolution. They are very bright
and their peak luminosities are relatively uniform.
Furthermore, the variations in peak luminosity that do exist
are related to
the width of their luminosity peak (hereafter this relation will
be referred to as the luminosity-width relation, or LWR). This
relation has been both calibrated by estimating the distances
to the host galaxies of nearby SNe Ia and simulated by
performing radiation transport upon models purported to span the
SN Ia event (H\"{o}flich \& Khokhlov 1996, Pinto \& Eastman 2000a, 2000b).
The combination of an extremely 
bright luminosity peak and the relatively well-determined
value of that peaks (via the empirical LWR) have permitted SNe Ia to be used as high-$Z$
distance indicators. Indeed, SNe Ia have been instrumental in
establishing that the Hubble Constant (H$_{0}$) has a value of
 $\sim$~70 km s$^{-1}$ Mpc$^{-1}$
(Freedman et al. 2001, Gibson et al. 2001) and in suggesting that the
cosmological constant has a non-zero value
(Perlmutter et al. 1999, Riess et al. 1998). These
uses of SNe Ia proceed despite controversy as to the exact
nature of SN Ia explosions.

Studies of the gamma-ray line emission from supernovae have long been
recognized as a powerful way to probe the nucleosynthesis and
explosion kinematics of these events (Clayton, Colgate \& Fishman
1969, Ambwani \& Sutherland 1988, Chan \& Lingenfelter 1991). The
$^{56}$Ni~$\rightarrow$~$^{56}$Co~$\rightarrow$~$^{56}$Fe decay chain
provides the most promising candidates for gamma-ray line studies of
prompt emission from SNe, producing strong lines at 158, 812, 847 and
1238 keV. At early times, the line fluxes increase as the expanding
ejecta unveils the radio-isotopes responsible for each line. The
timing of this unveiling is a function of both the distribution of the
isotopes and the kinematics of the ejecta. At later times, when the
ejecta asymptotically approaches being optically thin to the
gamma-rays, the line fluxes follow the isotopes' decay curves and
reveal the total production of each isotope.  Neither of these line
flux comparisons require an instrument capable of resolving the line.
If the line can be resolved, measuring the line profiles of the
individual lines shows the distribution (in velocity space) of the
radio-isotopes in the ejecta, allowing a very precise probe of the
nucleosynthesis in the SN explosion.  Because of their low mass, 
thermonuclear supernovae (type Ia SNe) have very strong gamma-ray 
signals and gamma-rays make ideal probes of the type Ia SN mechanism.

As the shape of the light curve peak (important in calibrating SN Ia's
for use as cosmological probes) is a direct function of the $^{56}$Ni
decay chain, determining the distribution of $^{56}$Ni in SNe Ia is a
primary goal of SN Ia studies.  At the elemental level, the same
diagnostics used in gamma-ray line studies are also present in the
optical and infra-red emission, and most current studies concentrate
upon that wavelength range.  However, optical and infra-red studies
require a much more in-depth knowledge of the ejecta characteristics
and suffer due to uncertainties in this knowledge.  Gamma-ray emission
studies have a number of features that allow a direct interpretation
of the observations and a more exact estimation of the $^{56}$Ni
yield.  The prompt emission lines from gamma-rays rely upon
the production of one isotope ($^{56}$Ni) and the determined
abundances do not suffer from line blends of a number of comparable
isotopes as they do in the optical or infra-red.  In addition, the
dominant opacity for gamma-rays in the energy range of most gamma-ray
lines is Compton scattering which varies smoothly with wavelength and
depends only weakly on the composition.  The gamma-ray emission lines
are produced by the decay of $^{56}$Ni and its decay product $^{56}$Co
which are insensitive to the ionization state of these isotopes.  In
contrast, the opacity in the optical and infra-red are dominated by a
complicated combination of the line opacities from all the elements in
the ejecta.  Beyond the difficulty of merely including this forest of
lines, the line opacities will depend sensitively both on the
composition and ionization state of the ejecta.  Gamma-ray emission is
a much more straightforward, and ultimately more accurate, probe of
the $^{56}$Ni yield in supernovae.

Despite the promise of studying prompt emission, only three SNe Ia
have even been observed with gamma-ray telescopes, resulting in only a
single, weak detection (SN~1991T) and two upper limits (SNe~1986G and
1998bu).\footnote{For this work, we distinguish ``prompt" emission
($^{56}$Ni \& $^{56}$Co decays) from ``SNR" and/or diffuse emission 
(such as $^{44}$Ti, $^{26}$Al and $^{60}$Fe decays).} 
In fact, although the
probability to detect prompt emission is predicted to be far higher
for thermonuclear SNe than for core-collapse SNe, the two strongest
detections have been from a SN of the latter type. SN 1987A was
detected at 847 and 1238 keV (from $^{56}$Co decays) by the SMM
instrument (Matz et al. 1989) and with balloon-borne instruments 
(Mahoney et al. 1988, Rester et al. 1989, Teegarden et al. 1989, 
Tueller et al. 1990, Kazaryan et al. 1990, Ait-Ouamer et al. 1990), 
 and at 122 keV (from $^{57}$Co decays) by CGRO's 
OSSE instrument (Kurfess et al. 1992).

Even so, the number of gamma-ray transport codes used in the
literature to study SNe Ia exemplifies the importance of these
diagnostics.  Preliminary comparisons between these simulations reveal
that the predicted fluxes vary considerably. Indeed, the variation
caused by different codes was larger than the variation caused by
different explosion models with the same code.  Disentangling the
differences between codes has been complicated by the fact that
the work in the literature does not use the same set of supernova
explosion models.  In addition, most of the published work is limited
to line fluxes, and different authors use different definitions of the
line flux (i.e.  whether the line flux results from the escape
fraction of the photons in the specified line or from the convolution
of an assumed instrument response over a simulated spectrum).  In this
paper, we eliminate the earlier confusion 
by directly comparing seven of the major codes used for 
gamma-ray line transport, using the same initial progenitors.

1) M\"{u}ller, H\"{o}flich \& Khokhlov (1991) and H\"{o}flich,
Khokhlov \& M\"{u}ller (1992) simulated emission from delayed
detonation models in anticipation of CGRO observations of SN 1991T.
That on-going effort, utilizing the MC-GAMMA code, produced a
number of papers, many of which studied the energy deposition in SN
ejecta.  A comprehensive paper by H\"{o}flich, Wheeler \& Khokhlov
(1998) explored various aspects of gamma-ray line emission, including
displaying spectra, line fluxes, line ratios and line profiles for
nine SN Ia models. More recently, they have also explored potential
ramifications of asymmetry upon the line fluxes and line profiles of
SN Ia emission (H\"{o}flich 2002).  We include this code in our
study, referring to it as ``H\"{o}flich".
%Hoflich, New Astronomy Reviews, 46, 475 (2002)

2) Shigeyama, Kumagai, Yamaoka, Nomoto, \& Thielemann (1991) simulated
gamma-ray emission for two SN Ia models, including the 1991T model,
W7DT.  Kumagai followed up that work by simulating more models
(including HECD) and treating the hard X/$\gamma$-ray emission from
their models (Kumagai \& Nomoto 1997, Kumagai, Iwabuchi \& Nomoto
1999), and more recently by studying the supernova contribution to the
cosmic gamma-ray background (Iwabuchi \& Kumagai 2001).  We include
this code in our study, referring to it as ``Kumagai".
%Shigeyama, T., Kumagai, S., Yamaoka, H., Nomoto, K., Thielemann, F.-K., 
%A\&AS, 97, 223 (1003)
%Iwabuchi, K., Kumagai, PASJ, 53, 669 (2001)
%Kumagai, S., Nomoto, K., In Ruiz-Lapuente, P., Canal, R., Isern, J. (Eds), 
%Proc. of the NATO ASI on Thermonuclear SNe (C486), Kluwer, Dordrecht, p515
%Kumagai, Iwabuchi \& Nomoto, Proc. of Astronomy with Radioactivities III, 
%eds R. Diehl, & D. Hartmann, (1999), p181

3) Other simulation efforts have been motivated by predicted
performances of specific missions and/or studies of energy deposition
in SN ejecta.  Burrows \& The (1990) studied X/$\gamma$-ray emission
from SNe Ia in anticipation of the launch of the COMPTEL and OSSE
instruments on the Compton Gamma Ray Observatory (CGRO), following
earlier, similar studies of SN 1987A (Bussard, Burrows \& The 1989,
The, Burrows \& Bussard 1990). That work investigated the energy
deposition in SNe (The, Bridgman, \& Clayton 1994), as well as the
supernova contribution to the cosmic gamma-ray background (The,
Leising \& Clayton 1993).  Milne, The, \& Kroeger (2000) and Milne,
Kroeger, Kurfess \& The (2002) used simulated gamma-ray line fluxes of
SN Ia models from this code to predict the performance of an
Advanced Compton Telescope.  We include this code in our study,
referring to it as ``The".
%Burrows & The, ApJ,360, 626 (1990)
%The, L.-S., Leising, M.D., Clayton, D.D., ApJ, 403, 32 (1993)
%Bussard,R.W., Burrows, A. & The, ApJ, 341, 401 (1989)
%The, Bridgman, W.T., Leising, Clayton, ApJS, 93, 531 (1994)
%Milne, Kroeger, R.A., Kurfess, J.D., The, L.-S., New Astronomy Reviews, 46, 617
%Milne, The, Kroeger, Proc. of 2nd Chicago Meeting on Thermonuclear Explosions, 
% eds Neimeyer, J., in press, astro-ph 0012078 (2000)

4) Isern, Gomez-Gomar \& Bravo (1996), Isern, Gomez-Gomar, Bravo \& Jean
(1997), and Gomez-Gomar,Isern \& Jean (1998) all displayed results of
an on-going study of gamma-ray emission from a range of SN Ia models.
Those studies have concentrated upon the potential for the INTEGRAL
satellite to detect that emission.  We include this code in our
study, referring to it as ``Isern".
%Isern, J., Gomez-Gomar, J., Bravo, E., 

5,6) Pinto, Eastman \& Rogers (2001) employed the FASTGAM code to
study the X/$\gamma$-ray emission from Chandrasekhar versus
sub-Chandrasekhar mass models of SNe Ia. This code was first developed
to study emission from SN 1987A (Pinto \& Woosley 1988).  The 3D
Maverick code (Hungerford, Fryer \& Warren 2003) was developed to
study asymmetries in core-collapse supernovae.  The physical processes
included in Maverick were chosen to match those in FASTGAM, though the
implementation techniques of these processes differed at the detailed
level.  We include the FASTGAM and Maverick codes in our study,
referring to them as ``Pinto" and ``Hungerford" respectively.
%Pinto, P.A., Eastman, R.G., Rogers, T., ApJ, 551, 231 (2001)
%Pinto, P.A., Eastman, R.G., ApJ, 530, 744 (2000a)
%Pinto, P.A., Eastman, R.G., ApJ, 530, 757 (2000b)  
%Woosley, S.E., Hartmann, D., Pinto, P.A., ApJ, 346, 395
%Hungerford, Fryer \& Warren, ApJ, 

7) In support of an effort to develop two next generation gamma-ray
telescopes, an Advanced Compton Telescope (Boggs \& Jean 2001) and a
High-Resolution Spectroscopic Imager (Harrison et al. 2003), Boggs
($ApJ$ submitted) simulated line profiles for SN Ia models.  
We include this code in our study, referring to it as ``Boggs".
% Boggs, S.E., Jean, P., A\&A, 376, 1126 (2001)
%Harrison, Fiona A.; Boggs, Steven E.;
% Christensen, Finn E.; Gehrels, Neil A.;
% Grindlay, Jonathan E.; Chen, C. M. H.;
% Craig, William W.; Hailey, Charles J.;
% Pinto, Philip; Thorsett, Steven;SPIE, 4851, 345 (2003)

In Section 2, we introduce the simulation techniques employed by the
seven groups and compare the physics that went into them.  In Section
3, we compare the spectral results from all codes and describe what
alterations were made to the codes to achieve agreement between all
groups.  In Section 4, we run a single code (The) with various models
and compare the different results possible for different explosion
scenarios. In Section 5, we compare these simulated spectra, 
with the SMM observations of
SN~1986G and the COMPTEL \& OSSE observations of SNe~1991T and 1998bu.

\section{Decay and Transport Physics}

\begin{figure}
\epsscale{1.0}
\plotone{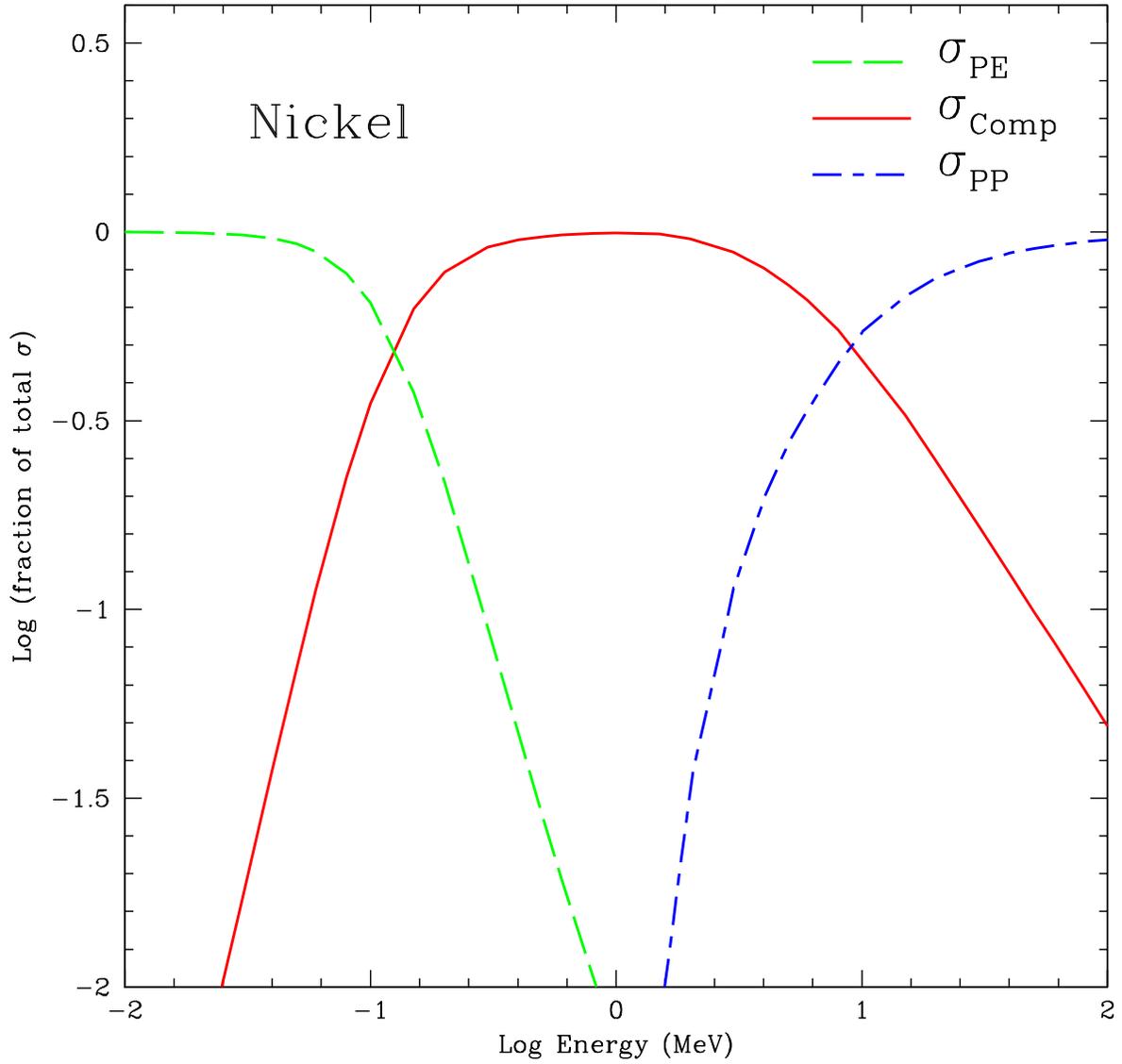}
\caption{Cross sections for photon interactions in nickel. 
Compton scattering (solid line) dominates over photo-electric 
absorption (dashed line) and pair-production (dot-dashed line) 
over the energy range 0.1 - 10 MeV.}
\label{cross}
\end{figure}

In order to understand the differences between the simulation
techniques used by the various groups in this comparison, we must
first lay out the basic physical picture of the problem we are 
representing numerically.  For the purposes of determining the high
energy spectrum (roughly 30~keV to 4~MeV) at the epochs of interest
(10 to 150 days), the assumption of a homologously expanding supernova
ejecta is valid.  The ejecta composition includes radioactive species
such as $^{56}$Ni and $^{56}$Co, the decay of which provides the
$\gamma$-ray line photons that generate the line and, through scatter
interactions, continuous $\gamma$-ray spectrum.  The basic interaction
processes involving these photons are pair production (PP),
photo-electric (PE) absorption, and Compton scattering off free and
bound electrons.  Figure \ref{cross} shows a plot of cross sections for these
interactions as a function of energy, which shows that the absorptive
opacities, PE and PP, are only dominant at low ( $< 150$~keV) and high (
$> 10$ MeV) energies, respectively.  The majority of the energy range
discussed is dominated by Compton scattering interactions.

As discussed in detail by Ambwani \& Sutherland (1988), the picture
described above is well-suited for Monte Carlo transport
methods.  Since six of the seven groups in our
collaboration have employed this technique (H\"oflich, Hungerford,
Isern, Kumagai, Pinto and The), we briefly recap the
major points described by Ambwani \& Sutherland (1988).  The
fundamental advantage of Monte Carlo is its ability to accommodate very
complicated physical processes in the transport.  This is accomplished
by simulating the {\it micro}-physics of the photon's propagation
through the supernova ejecta.  The principle is very straightforward:
the mass of nickel atoms in the input model implies a certain amount
of radioactive decay luminosity.  Monte Carlo packets 
(which represent some quantum of photon luminosity) 
are then launched in proportion to the decay rate and 
the mass distribution of nickel atoms.  Each packet's energy is chosen
in proportion to the branching ratios of the possible decay lines and
its initial direction is picked at random, assuming isotropic emission. 
The emitted packet is then allowed to propagate through the ejecta,
interacting with the material through scattering and absorption.  This is a
microscopic treatment of the transport in the sense that each individual
packet of photons is tracked through each individual interaction.

The likelihood of a photon experiencing an interaction during its
flight is dictated by the total cross section for interaction
($\sigma_{tot}$).  When an interaction occurs, the type of
interaction, scatter or absorption, is chosen randomly in proportion
to the ratio of $\sigma_{scat}/\sigma_{tot}$ or
$\sigma_{abs}/\sigma_{tot}$.  The well-described micro-physics of the
PP and PE absorption and the Compton scatter process are explicitly
taken into account for each packet interaction, and are thus treated
with no approximation.  When a packet's path brings it to the surface
of the ejecta, it is tallied into the escaping SN spectrum.  Likewise,
if the path ends in an absorption, the packet's energy is deposited
into the ejecta.  In this way, the Monte Carlo transport technique
allows for straightforward calculation of the emergent hard X- and
$\gamma$-ray spectrum, as well as energy deposition into the ejecta
via photon interactions.

If the emerging line profile of the $\gamma$-ray decay lines is the
only quantity of interest, semi-analytic techniques alone, as employed
by Boggs in this comparison, can be effectively used
as well.  The Compton equation describes the energy shift a photon
experiences upon suffering a Compton scatter.  For the decay lines we
are interested in (E$\sim$1~MeV), a single Compton scatter generally
shifts the photon's energy out of the decay line profile.  This means
that the line profiles in the emergent spectrum arise primarily 
from photons that escape the ejecta without any interaction, with a 
secondary contribution from forward-scattered Compton photons. 
The line profiles can thus be calculated analytically
by multiplying the emitted luminosity, as determined from the mass
distribution of radioactive species in the ejecta, by the factor
$e^{-\tau}$, where $\tau$ is the total optical depth from the
emission point to the surface of the ejecta.  Analytical techniques
such as this provide an invaluable test of the more computationally
intensive Monte Carlo technique described above.

Regardless of technique chosen, bringing the physical picture
to a numerical representation requires a series of
computational decisions.  In the following sub-sections we will review
the physics pertinent to these computational choices.  These 
choices fall into three primary categories:
\begin{itemize}
\item[2.1)]{Description of the Ejecta (Differential Velocity, 
Density Evolution)}
\item[2.2)]{Photon Source Parameters (Lifetimes and Branching Ratios, 
Positron Annihilation, Ejecta Effects, Weighting)}
\item[2.3)]{Opacities for Photon Interactions (Compton Scattering, 
Photo-Electric Absorption, Pair Production and Bremsstrahlung Emission)}
\end{itemize}
Table \ref{tab_sim} lists
the various codes and provides information regarding the
numerical implementations of the physics discussed below.

\subsection{Ejecta}

For the different explosion models, the ejecta is determined by
mapping the model into spherical Lagrangian mass zones and expanding
this ejecta homologously outward with time.  Taking snapshots in time
of this ejecta, each gamma-ray calculation uses the density,
radius, velocity and composition of the ejecta for these mass
zones\footnote{The 3-dimensional codes must first map the ejecta into
  a 3-dimensional grid.  The number and type of nuclei treated in each
  code varies slightly and abundances were interpolated to match each
  code separately.}.  Some codes simply take the position of the
$^{56}$Ni and $^{56}$Co, but others include the motion of the ejecta
at varying levels of sophistication. The two major velocity effects
are the differential motion and the density reduction due to
expansion.

\subsubsection{Differential Velocity}

\begin{figure}
\epsscale{1}
\plotone{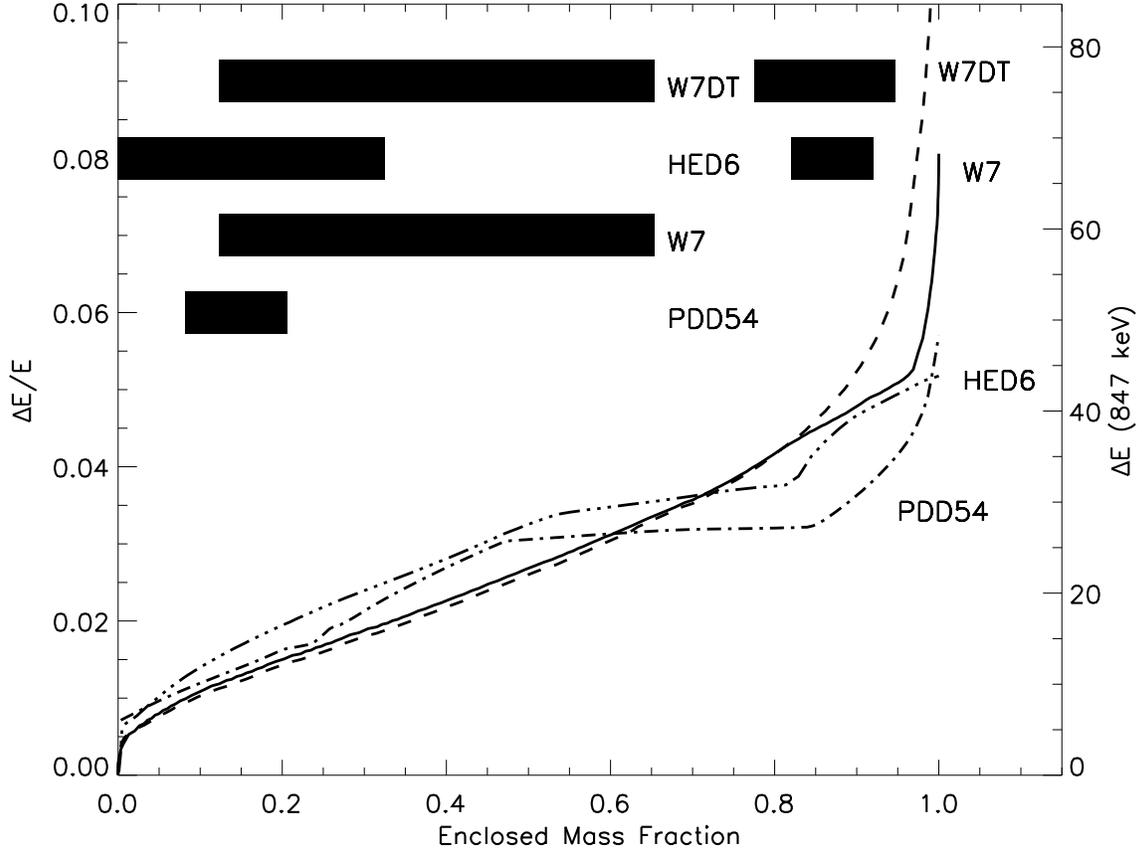}
\caption{Line shifting due to the expansion of the 
ejecta of four SN models. The fractional line shift due 
to the expansion of the ejecta is plotted on the 
left axis, the shift for the 847 keV line is shown 
on the right. For reference, $^{56}$Ni-rich regions 
of the ejecta are shown in the upper left as thick,  
horizontal bars.}
\label{broad}
\end{figure}

Since the radioisotope is distributed in velocity space and the 
opacity depends on the relative velocities, the ejecta velocity 
will affect the propagation of the photon packets.  
The packets are created with a decay line
energy in the co-moving frame of the surrounding ejecta, but are tallied
in the rest frame of the observer.  The Doppler shift between these 
two frames is the dominant source of broadening in the line profiles.
In Figure \ref{broad}, we show the amount of line broadening possible 
for four SN Ia models. 
In addition, as the packet propagates through the ejecta, its 
energy, as measured in the local co-moving ejecta frame, is constantly
changing.  Since interaction cross sections are energy-dependent, the
opacity through the ejecta for the packet will be different
from the case where ejecta velocity is neglected.  For our scenario, 
this is a small effect, as our dominant opacity (Compton
scattering) is a slowly varying function of energy.

The Boggs, H\"oflich, Hungerford, Isern and Pinto algorithms included the 
ejecta velocity effects, allowing them to calculate detailed line profiles
(Table 1). 

\subsubsection{Density Evolution} 

Assuming the decision was made to account for ejecta velocity effects,
one must then choose whether to allow this motion to feed back on the
densities throughout the ejecta.  The photon packet does not traverse 
its path infinitely quickly.  Indeed, there is some flight time 
associated with each packet trajectory, and during this flight
time, the ejecta undergoes expansion.  This results in lower
densities, and thus lower opacities, as the packet propagates through 
the star. The alternative to treating this expansion is to assume 
the transport takes place
within a differential time slice $dt$, over which the hydrodynamic
quantities do not evolve at all.  For a homologously expanding ejecta,
the density falls off simply as $t^{-3}$, making this feedback effect
easy to implement.  However, accounting for it is only a partial step
toward a time-dependent treatment of the problem.  The source of the
photon packets must also be treated in a time-dependent fashion in
order to be self-consistent.  Unfortunately, the implementation of the
source's time-dependence is not trivial in a Monte Carlo treatment.

Pinto allowed for the ejecta expansion to feedback on the densities.  
The semi-analytic technique employed by Boggs accounted for both the expansion 
feedback and the time dependence of the photon source (i.e. photons
from the far side of the ejecta take longer to arrive at the detector and 
must be launched at an earlier time during the explosion. See \S 2.2.3)

\subsection{Photon Source}

Differences in the gamma-ray sources include not only $^{56}$Ni and 
$^{56}$Co decay times and branching ratios, but the emission from 
positron annihilation.  The actual photon emission also depends on 
the ejecta.  Finally, the method of weighting the packets can also 
pose a problem when normalizing the escaped packet counts into 
physical flux units.

\subsubsection{Decay Times and Branching Ratios}

The source of photons for these high energy calculations is
exclusively $\gamma$-ray line emission from the decay of various
radio-isotopes present in the supernova ejecta.  The fundamental decay
chain is that of the radio-isotope $^{56}$Ni.  The SN explosion
synthesizes $^{56}$Ni, which promptly decays via electron capture to
$^{56}$Co with a mean lifetime of $\sim$8.8 days.  The $^{56}$Co
produced in this decay is also unstable, though with a longer
lifetime ($\sim$111.4 days).\footnote{We show in Table 
\ref{tab_dec} half-lives from the Nuclear Data Sheets (Junde 1999) 
and branching ratios from the 8th edition of the Table of Isotopes 
(Firestone \& Shirley 1996). It is apparent from the lower portion 
of Table \ref{tab_dec} that earlier versions of these tables (and other tables 
such as ``Table of Radioactive Isotopes", Browne \& Firestone 1986) 
contained lifetimes that were as long as 113.7 days mean lifetime for 
$^{56}$Co and as short as 8.5 days mean lifetime for $^{56}$Ni. This has lead to 
confusion in the literature as to the correct values.} However, 
we expect the errors caused by the decay times to be less than $\sim$5\% 
(Figure \ref{decrs}).

Whereas the $^{56}$Ni decay always proceeds via electron capture, the
$^{56}$Co decay proceeds either through electron capture (about 81\%
of decays) or positron production 
(roughly 19\% of decays).\footnote{It has been suggested (Mochizuki et al. 1999) 
that the ionization state of the gas can affect the electron-capture decay rates 
in supernova remnants, since these decays ($^{56}$Ni, $^{56}$Co, $^{44}$Ti) 
proceed mainly by capturing inner-shell electrons.
This effect cannot be important in the pre-remnant phase, those times before
shocks with the circumstellar material have heated the gas to millions 
of degrees.  The gas temperature in the supernova at times considered in this work is 
always far too low to for inner shells to have a significant vacancy probability. 
Further, the timescale over which atoms with an inner shell vacancy due to 
non-thermal ionization fill that shell by relaxation from outer shells is far smaller 
than the mean time between ionizations. The decay rates are thus essentially the 
zero-ionization (laboratory) values, and these are the values we have employed.} 
Shown in Table \ref{tab_dec} are the relative abundances of the dominant lines
from the $^{56}$Ni and $^{56}$Co decays.  Note that these values refer
to the number of photons emitted per 100 decays of the respective
isotope (i.e.  this includes the effects of the 19\% positron
production branching ratio).  Clearly, the dominant branches, both for
studies of gamma-ray line emission and for studies of the energy
deposition are the 158, 812, 847 and 1238 keV lines.  The exact values
for branching ratios and lifetimes of these radioactive decays are
subject to updates and revisions, as one might expect.  As a result,
the suite of values used in a $\gamma$-ray transport code are chosen
from a range of possibilities available in the refereed literature.

For the most part, the values adopted from different references have
no noticeable affect on the calculated spectra.  The only significant
variations in adopted branching ratios from earlier works to the
current simulations was with the H\"oflich code. In previous works,
the H\"oflich code adopted 0.74 for the 812~keV line of $^{56}$Ni
decay rather than the 0.86 employed by the other groups.  Also, in
previous simulations with the H\"oflich code, it was assumed that the
positron production branch left the $^{56}$Fe daughter nucleus always
in its ground state (M\"{u}eller, H\"oflich, Khokhlov 1991).  This led
to branching ratios for the $\gamma$-ray lines from excited $^{56}$Fe
being reduced from the published values by the 19\% positron
production branching ratio.  

\subsubsection{Positron Decay}

Absent from Table \ref{tab_dec} are the 511 keV line and positronium
continuum which result from the positron production branch of the
$^{56}$Co decay.  These positrons are created with $\sim$~600 keV of
kinetic energy that must be transferred to the ejecta before
the positron can annihilate with electrons in the ejecta. It is
usually assumed that during the epoch of interest for gamma-ray line
studies ($\leq$ 150 days), positrons thermalize quickly and thus have
negligible lifetimes, annihilating {\it in-situ}.  Detailed positron
transport simulations (Milne, The \& Leising 1999) have shown
that this is not a wholly correct assumption at 150 days; however,
only a small error is introduced by making this assumption. Although
it is reasonable to assume that the positrons annihilate promptly,
in-situ, the nature of the resulting emission is not clear. Depending
upon the composition and ionization state of the annihilation medium,
the positron can annihilate directly with an electron (and produce two
511 keV line photons in the rest frame of the annihilation), or it can
form positronium first. If positronium is formed (and the densities
are low enough to not disrupt the positronium atom), 25\% of the
annihilations occur from the singlet state.  Singlet annihilation
gives rise to two 511 keV line photons, as with direct
annihilation. However, 75\% of annihilations occur from the triplet
state, which gives rise to three photons. As the three photons share
the 1022 keV of annihilation energy, a continuum is produced. This
continuum increases in intensity up to 511 keV and abruptly falls to
zero.

The resulting spectrum can thus be characterized by the positronium fraction, 
f(Ps), 
a numerical representation of the fraction of annihilations that form 
positronium (e.g. Brown \& Leventhal 1987):

\begin{equation}
f(Ps) = \frac{2.0}{1.5 + 2.25(A_{511}/A_{posit})},
\end{equation}
\label{eq_fps}

\noindent
where $A_{511}$ and $A_{posit}$ are the observed 511 keV line and the
positronium three-photon continuum intensities, respectively.
Positronium fractions range between 0 - 1, with most researchers
assuming that SN annihilations have a similar positronium fraction as
the Galaxy.\footnote{Note that the positronium fraction function
  cannot accept continuum fluxes of 
exactly zero. If $A_{posit}$ = 0.0, then f(Ps) = 0.0, independent of the 
equation.} Utilizing wide-FoV TGRS observations of galactic positron 
annihilation, Harris et al. (1998) estimated the positronium fraction to be 
0.94$\pm$0.04.  Similarly, utilizing CGRO/OSSE observations of 
the inner Galaxy, Kinzer et al. (2001) estimated the positronium fraction to 
be 0.93$\pm$0.04, both values in agreement with theoretical estimates of
interstellar medium (ISM) positron annihilation.  However, the composition of
SN Ia ejecta is far different than the ISM, being dominated by
intermediate and heavy elements rather than hydrogen and helium. Thus,
ISM annihilation is completely different than SN ejecta annihilation.
Likewise, the galactic annihilation radiation measured by OSSE is a
diffuse emission, and thus it is distinct from the in-situ annihilations
that occur in SN Ia ejecta within 200 days of the SN explosion.  The
expectation is that charge exchange with the bound electrons of these
intermediate and heavy elements would lead to SN ejecta having a
positronium fraction of at least 0.95. However, a zero positronium
fraction for annihlations that occur in SN ejecta cannot be ruled out.

For our purposes here, it suffices to say that the expected spectrum
from positron annihilation is uncertain, and the individual members of
this comparison team have adopted positronium fractions of either 0
(Hungerford, Kumagai, and Pinto) or 1 (H\"oflich, Isern, and The); see
Table \ref{tab_sim} for a summary.  The three groups employing
positronium fractions of 1 adopted the energy distribution of the 
positronium continuum treatment in Ore \& Powell (1949).  

\subsubsection{Ejecta Effects on Decay}

\begin{figure}
\epsscale{1.0}
\plotone{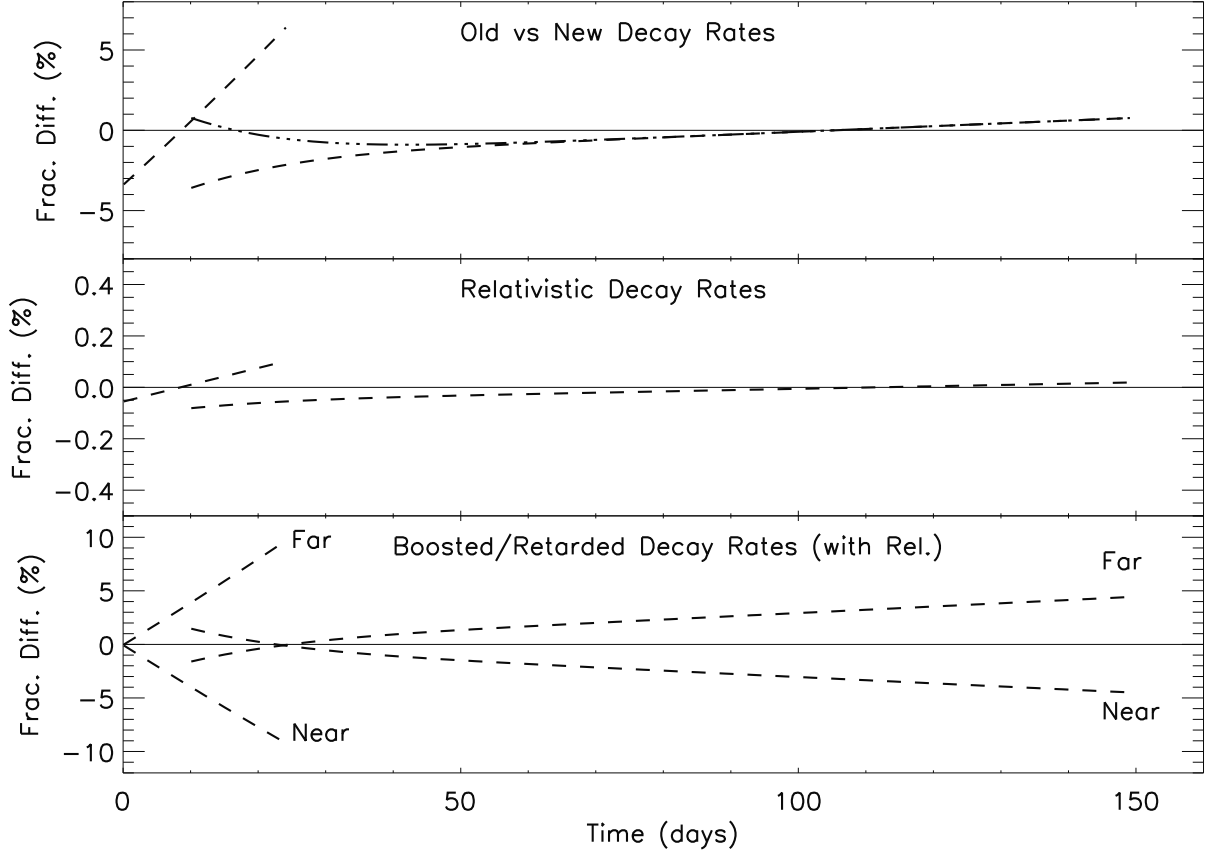}
\caption{Nickel and Cobalt decay rates. The upper 
panel shows the fraction change in the decay rates 
for $^{56}$Ni and $^{56}$Co assuming 8.8$^{d}$ and 
113.7$^{d}$ mean lifetimes rather than 8.5$^{d}$ and
111.5$^{d}$ (dashed lines). The simplified $^{56}$Co decay rate 
used in HWK98 compared with the 8.5$^{d}$ and 111.5$^{d}$ lifetimes 
is also shown (dot-dot-dot-dashed line). 
The middle panel shows the fractional change in the 
decay rates produced by considering the relativistic effects of 
the ejecta's expansion velocity upon the decay rates. 
The lower panel shows the effect of the boosting/retarding of 
the decay rate to synchronize all photons to arrive simultaneously 
with photons from the center of the SN ejecta. 
The ejecta velocity is assumed to be 10,000 km/s in the lower two 
panels, a relatively large value for  $^{56}$Ni-rich ejecta. Bear in 
mind that until the ejecta becomes thin to gamma-rays, the 
emission from the decays near the surface on the 
near edge will dominate the integrated emission.}
\label{decrs}
\end{figure}

The motion of the ejecta can also change the decay rate.  The decay
equations for $^{56}$Ni and $^{56}$Co decays in a stationary medium
are:

\begin{equation}
\frac{1}{{\mathrm{Ni}}_{o}} \left( \frac{dNi}{dt} \right) = -
\frac{1}{\tau_{Ni}}exp \left( \frac{-t}{\tau_{Ni}} \right ),
\end{equation}
\label{eq_decni}

\begin{equation}
\frac{1}{{\mathrm{Ni}}_{o}} \left ( \frac{dCo}{dt} \right ) =
\frac{-1}{\tau_{Co} -\tau_{Ni}} \left[ exp \left( \frac{-t}{\tau_{Co}}
  \right) -exp \left( \frac{-t}{\tau_{Ni}} \right) \right],
\end{equation}
\label{eq_decco}

\noindent
where, $\tau_{Ni}$ and $\tau_{Co}$ are the mean lifetimes of the
isotopes, Ni$_{o}$ is the $^{56}$Ni produced in the SN explosion, and
$t$ is the time since explosion.  For a given model time ($t=t_m$),
these equations can be solved for the number of Nickel and Cobalt
atoms that will decay during an infinitesimal time slice $dt$.  These
equations still hold for a finite time step $\Delta t$, assuming
$\Delta t$ is much less than the lifetime $\tau$.  Strictly speaking,
the lifetimes ($\tau_{Ni}$ and $\tau_{Co}$) in the above equations,
are in the frame of the isotope, which is moving relative to an
external observer.  Since the velocity of the ejecta can be upwards of
10,000 km~s$^{-1}$, an exact treatment of the decay rate must include
a conversion to the frame of the external observer.  This relativistic 
effect is proportional to $\gamma=(1-v^2/c^2)^{-1/2}$ and is only 
a 0.1-0.2\% effect overall (Fig. \ref{decrs}).  Aside from Boggs, none of the 
codes include this effect.

More important is the flight time of the photons through the ejecta.
In the context of Equations (2) and (3) it is straightforward to point
out where to accomplish this.  Emission from the near side of the
ejecta should be calculated from the above equations using a retarded
time relative to the far side.  In this way, photons from the front
and back of the ejecta arrive simultaneously at the detector.  Figure
\ref{decrs} shows the effect these two issues (in the extreme) 
have on the calculated decay rate.  The flight time of the
photons introduces less than a 10\% error.  (N.B. The dash-dot-dot line
in Figure~\ref{decrs} represents the variation in decay rate of Cobalt
resulting from an approximate form of Equation 3 used in one of the
comparison codes.)  Again, Boggs' code is the only one that
incorporates these effects.

\subsubsection{Weighting}

The last uncertainty is purely numerical in nature and arises from 
the weighting (and subsequent normalization) of the photon packets.
Combining the decay rate with the branching ratios, which provide a
measure of the average number of photons per decay, Equations (2) and
(3) yield a total photon luminosity (${\mathcal L}_{phot}$) of the
ejecta (in phot/s).  Given the number of photon packets to be tracked
in the simulation (${\mathcal N}_{packet}$), the weight of each packet
is

\begin{displaymath}
{\mathcal W}_{packet} = \frac{{\mathcal L}_{phot}}{{\mathcal N}_{packet}}.
\end{displaymath}

\noindent
More complicated weighting algorithms are possible, and provide
advantages when specialized information is desired.  For example,
detailed studies of the spectral characteristics for weaker decay
lines benefit from emitting a large number of packets at the 
decay energies of interest.  In this way, the signal-to-noise of the
spectrum at those weak lines is enhanced beyond what the uniform
weighting technique could provide.  In any case, the normalization
applied via this weight factor can be taken into account from within
the transport code itself, or as a post-process step on the photon
packet counts, which result from the base Monte Carlo transport
routine.  The validity of the normalization is easily tested through
the analysis of the integrated line flux lightcurves for the various
decay lines.  These lightcurves can be directly compared with the
semi-analytic technique discussed above for decay lines with energies
greater than about 1~MeV (i.e. where the continuum has a negligible
contribution to the spectrum.)  For our study, all the Monte Carlo
algorithms were run using constant weight packets to reduce the
complexity of the comparison, but as we shall see, it is the
weighting and the subsequent normalization of the flux that caused
many of the discrepancies in past simulations (see \S 3.4).

\subsection{Photon Interaction Processes}

Once the decay photons have been created, their propagation through
the ejecta is dictated by the three interaction processes mentioned at
the start of this section: Pair-Production, Photo-Electric absorption
and Compton scattering.  The major features of the spectrum, with the 
exception of actual line fluxes, can be understood primarily through
the PE absorption and Compton scatter interactions.

\subsubsection{Compton Scattering}

For the majority of the energy range we are interested in, the Compton
scatter interaction off bound and free electrons dominates.  This
interaction depends only on the total electron density in the ejecta
and energy of the incident photon.  Since almost all SN Ia ejecta has
an electron fraction Y$_{e}$~$\sim$~0.5, this interaction is only
weakly dependent upon the composition.

Figure \ref{cross} shows the
energy dependence of the cross-section for Compton
scattering as employed by the various groups. This cross-section
is a smoothly varying function of energy and, in general, is
represented

\begin{equation}
\sigma_{\rm Compton} = \frac{3\sigma_{Th}}{8\epsilon} \left[ \left[
1 - \frac{2(\epsilon + 1)}{\epsilon^{2}} \right] {\mathrm ln}(2\epsilon +1)
+\frac{1}{2} + \frac{4}{\epsilon} - \frac{1}{2(2\epsilon +1)^2}
\right],
\end{equation}

\noindent
where $\sigma_{Th}$ is the Thomson scattering cross-section, and
$\epsilon$ is the ratio of the photon energy to the electron rest
mass.

While photo-electric absorption and pair-production interactions consume the 
photon, the scattering process produces a lower energy photon traveling
in a new direction.  The down-conversion of the photon's energy is 
the dominant process for populating the hard X-ray continuum, and the 
exact energy distribution of the outgoing photons is described by 
the Klein-Nishina (KN) differential scatter cross-section.  The KN 
formula is given by (Raeside, 1976)

\begin{equation}
\frac{d\sigma}{d\epsilon'} = -\frac{3\sigma_{Th}}{8}  
\left( \frac{1}{\epsilon} \right)^2 \left[ \frac{\epsilon}{\epsilon'} +
\frac{\epsilon'}{\epsilon} - 1 + \left( 1 - \frac{1}{\epsilon'} +
\frac{1}{\epsilon} \right)^2 \right], 
\end{equation}

\noindent
where $\epsilon$ is the photon's incoming energy and $\epsilon'$ is
the photon's outgoing energy.  Given $\epsilon$, many techniques exist
for sampling an outgoing energy from this relation.  Combining this
information with the Compton formula, an outgoing photon direction is
then determined.  Detailed comparisons of the individual
sampling techniques used by the various groups have not been done.
However, for the six groups that track the scattered photons, the
continuum in their simulations is produced entirely through the
scatter interaction.  Fortunately, the shape of this Comptonized
continuum (200~keV - 800~keV) is a direct and sensitive test that the
physics of photon-electron scattering has been implemented appropriately.

\subsubsection{Pair-Production and Photo-Electric Absorption Opacities}

At low energies (less than $\sim$~200~keV), the smooth, nearly power
law continuum created from Compton scattering suffers a turn over due
to photo-electric absorption effects.  Just as in the adoption of
values for branching ratios and decay lifetimes, the literature offers
more than one reference for choosing absorptive opacities.  The PE and
PP opacities employed by the various groups in our collaboration can
be found from three primary references (Viegle, Hubbell and ENDL), 
which provide these cross sections in tabular form
(by energy and proton number.)  Techniques for interpolating cross
sections from the provided energy table values varied among the
different groups.  The number of nuclei species (different proton
numbers), which were considered as contributors to these absorptive
opacities, were also treated differently in the various codes.  These
types of variations in the numerical implementation ought to manifest
themselves as slight changes in the location of the low energy
spectral cut off.

In addition, both of these absorptive interactions allow for the
possibility of high energy photon daughter products: annihilation
photons for the case of pair production, and X-ray fluorescence
photons for the case of photo-electric absorption.  The decision
to include these processes and the technique for implementing
them varied among groups.  The X-ray fluorescence photons are 
below the low energy cutoff and, thus, contribute predominantly to the 
calculated deposition energy.  In this paper we concentrate only on the
emergent spectrum, and thus do not probe the differences caused
by the inclusion of the X-ray fluorescence.

\subsubsection{Bremsstrahlung Emission}

Another important photon emission process from the ejecta is the
bremsstrahlung process of the energetic Compton-recoil electrons (E
$\leq$ 3 MeV; recoiling from Compton scattering events with the
primary radioactivity gamma rays).  This bremsstrahlung process takes
place in all supernovae that are powered by radioactive decay.  The
large abundance of these electrons gives rise to the dominance of
bremsstrahlung photons as the hard X-ray source; i.e., below 30 keV
and 60 keV at 20$^{d}$ and 80$^{d}$, respectively in both models W7
and DD4 (Clayton \& The 1991; Pinto, Eastman, \& Rogers 2001).  
The shape of the bremsstrahlung spectrum emerging from the surface
is sensitive to the photoelectric opacity and with the flux,
$F(E) \propto E^{\alpha}$ 
where $\alpha$ is $\sim$1.1 and $\sim$1.8 at 20$^d$ and 80$^d$,
respectively for model W7 (Fig. 13 of Clayton \& The 1991);
the spectral luminosity increases slowly between 1 and 60 keV.
The sudden change in the hard X-ray slopes
between 10 keV and 100 keV (from the bremsstrahlung spectrum at lower
energies to the Compton scattering spectrum at higher energies) can be
used as the signature of this process.
None of the simulations in this 
comparison project included this process.

\section{Comparisons between Codes}

The seven codes included in this study have all
produced published simulations of SN models. All but the
Hungerford code have produced published simulations
of specifically SNe Ia. 
Indirect comparisons between published works from the codes being
studied in this paper suggest that different codes reach different
answers.  Notably, HWK98 and Kumagai \& Nomoto (1997) 
both predict larger line
fluxes than Pinto, Eastman \& Rogers (2001), Milne, Kroeger \& The
(2001) or Boggs ($ApJ$ submitted).  However, determining the cause of such
spectral variations has been difficult since no single input Ia model
has been simulated by all groups.  
While it is generally agreed that SNe Ia are caused by the
thermonuclear explosion of an accreting WD, there remains considerable
controversy as to the exact nature of the progenitor and the physics
behind the development of the burning front: deflagration
vs. detonation, number of ignition sites (e.g. Livio 2000).  These
differences have produced a set of SN Ia explosion models in terms of
a handful of parameters that form the basis for comparisons with SN
observations.
In this paper, we provide the much needed direct comparisons by 
running all seven gamma-ray transport codes
on the same set of SN~Ia explosion model inputs.  The set of three models that
were selected for comparisons are DD202C (a Chandrasekhar-mass delayed
detonation, H\"{o}flich, Wheeler \& Thielemann 1998), HED6 (a sub-luminous,
sub-Chandrasekhar mass Helium detonation, H\"{o}flich \& Khokhlov
1996) and W7 (a Chandrasekhar-mass deflagration, Nomoto, Thielemann \&
Yokoi 1984).  In Table \ref{tab_mods}, we show the relevant
characteristics of the models.  Errors were introduced by
imperfections in the conversion of each model into the varied formats
required by each code.  Typically these errors were 2-3\% of the
mass or kinetic energy and were found to have a negligible
effect upon the Compton-scattering dominated portion of the
spectra. 

For these comparisons, we focus on three aspects of the gamma-ray 
calculations:  
the overall spectra, the line profiles and, the most-important observed 
quantity in the near future, the line flux.

\subsection{Overall Spectra}

\begin{figure}
\epsscale{0.91}
%\plotone{ps/dd202c_15.ps}
%\plotone{ps/dd202c_25.ps} 
%\plotone{ps/dd202c_50.ps} 
\plotone{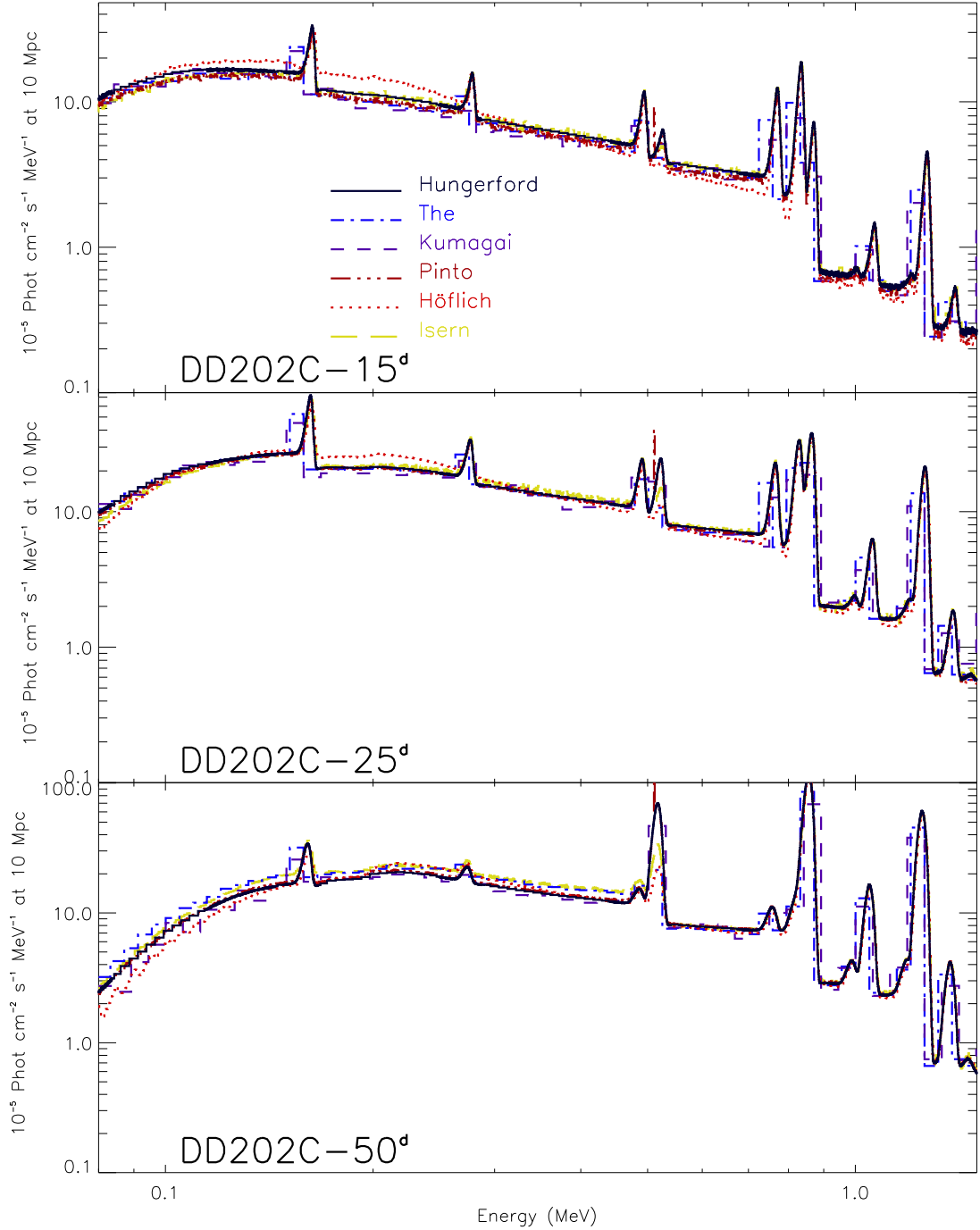}
\caption{A sequence of spectra for the SN Ia model, DD202C. 
The spectra, at 15$^{d}$, 25$^{d}$, and 50$^{d}$, show the 
level of agreement between simulations for both the 
line and continuum emission. Comparisons between the two 
algorithms that do not treat line broadening/shifting 
(The \& Kumagai) and the others that do, show the 
early effects of blue-shifting. }
\label{sp_dd202c}
\end{figure}

\begin{figure}
\epsscale{0.91}
%\plotone{ps/w7_15.ps}
%\plotone{ps/w7_25.ps}
%\plotone{ps/w7_50.ps}
\plotone{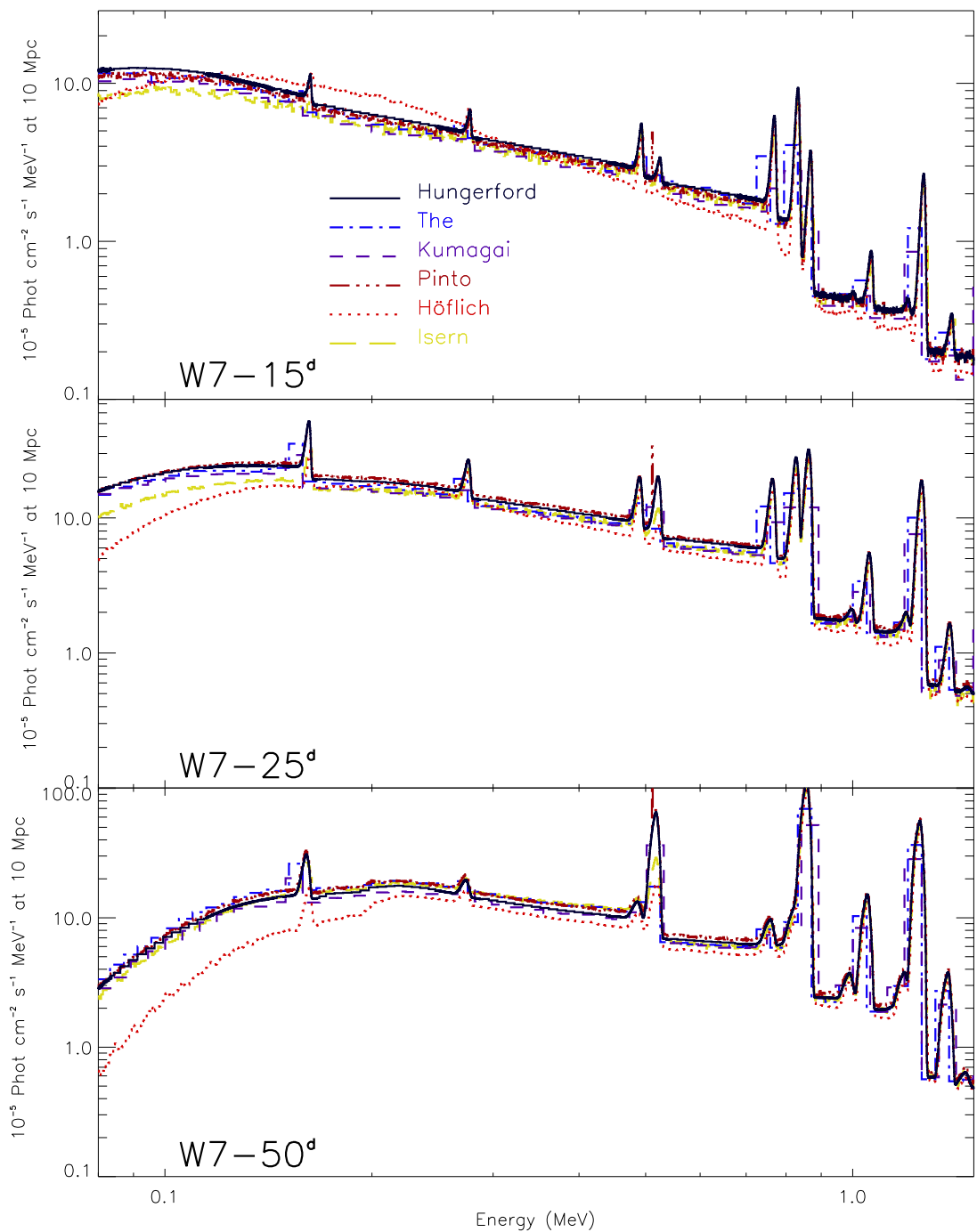} 
\caption{A sequence of spectra for the SN Ia model, W7.
The absence of nickel near the surface of W7 leads to 
the inhibition of line emission until later times. 
As with DD202C, the spectra, at 15$^{d}$, 25$^{d}$, 
and 50$^{d}$, show a high level of agreement between 
simulations for both the line and continuum emission.}
\label{sp_w7}
\end{figure}

\begin{figure}
\epsscale{0.91}
%\plotone{ps/hed6_15.ps}
%\plotone{ps/hed6_25.ps}
%\plotone{ps/hed6_50.ps}
\plotone{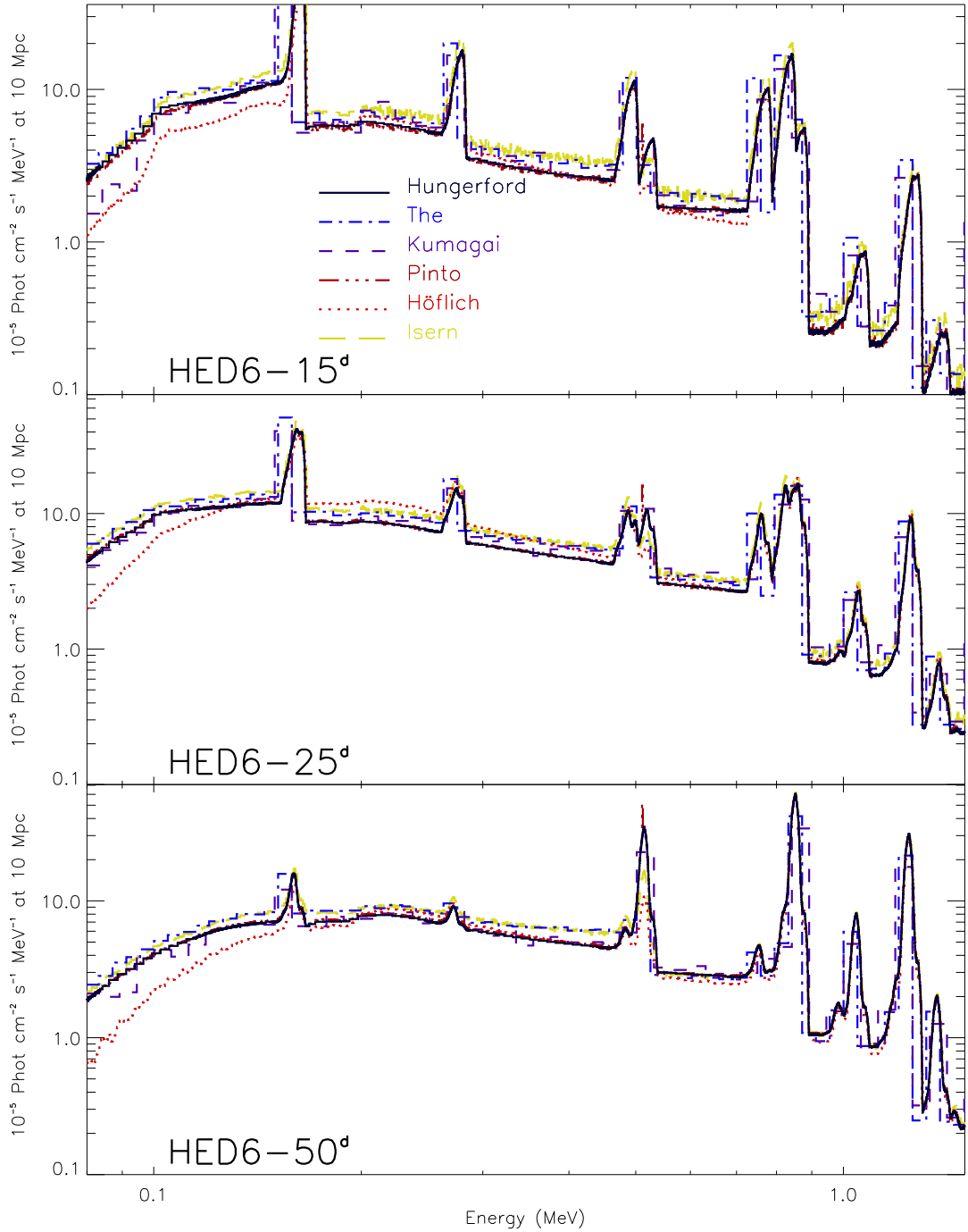}
\caption{A sequence of spectra for the SN Ia model, HED6.
The spectra, at 15$^{d}$, 25$^{d}$, and 50$^{d}$, show a 
high level of agreement between simulations, in this case 
for a low-mass model that features early escape of gamma-ray 
emission. }
\label{sp_hed6}
\end{figure}

Figures \ref{sp_dd202c} - \ref{sp_hed6} show a sequence of spectra
from simulations of DD202C, W7, and HED6, respectively.  These
spectral results arise from current versions of the 6 Monte Carlo
codes employed in this comparison and agree to within the statistical
noise except in a few cases.  $\S$3.3 describes in detail the
necessary corrections that were made to arrive at the current
versions.  The remaining differences in the spectral simulations can
be isolated in terms of the physical processes outlined in $\S$2.
For example, in Figures \ref{sp_dd202c} and \ref{sp_w7} at the
earliest epoch, it is clear that the H\"oflich spectra exhibit a
different continuum slope across the rough energy range of
200~keV~-~800~keV.  The shape of the continuum in this portion of the
spectrum is dictated primarily by the Klein-Nishina differential
scattering cross section, although physical effects such as Doppler
corrections for the ejecta velocities may also change the overall
spectral slope.  Closer inspection of the Compton scatter and Doppler
boost routines between H\"oflich and other codes did not reveal an
obvious cause for this difference, which has a maximum magnitude of
order 30\% but is much smaller across most of the energy range.

As discussed in $\S$2.2.2, spectral variations due to differences in
the assumed positronium fractions should appear in the 400 - 550 keV
energy range (Figures \ref{sp_dd202c} - \ref{sp_hed6}).  At late
times, one would expect the codes that include the positronium
continuum to have slightly higher continuum spectra and weaker
lines.  There is very little difference between the codes that include
a positronium continuum component (H\"{o}flich, Isern, and The) and
those that do not (Hungerford, Kumagai, and Pinto), but the expected
trends seem to hold.  As these spectra likely bracket the range of
possible annihilation spectral features, the treatment of the
positronium fraction primarily affects the strength of the 511 keV
line, and it does not dominate the appearance of the continuous
spectrum.

There also remain differences in the $\leq$ 100 keV spectra that
exceed statistical fluctuations.  These differences likely 
arise from differences in the implementation of 
photoelectric absorption opacities.  Differing interpolation
techniques for the tabular opacities, to account for the difference in 
number of nuclear species treated, may be responsible for 
these discrepancies.  As the emphasis of this comparison is on the 
higher energy gamma-ray portion of the spectrum, we did not attempt to 
resolve these opacity differences.

\subsection{Line Profiles}

\begin{figure}
\epsscale{1}
\plotone{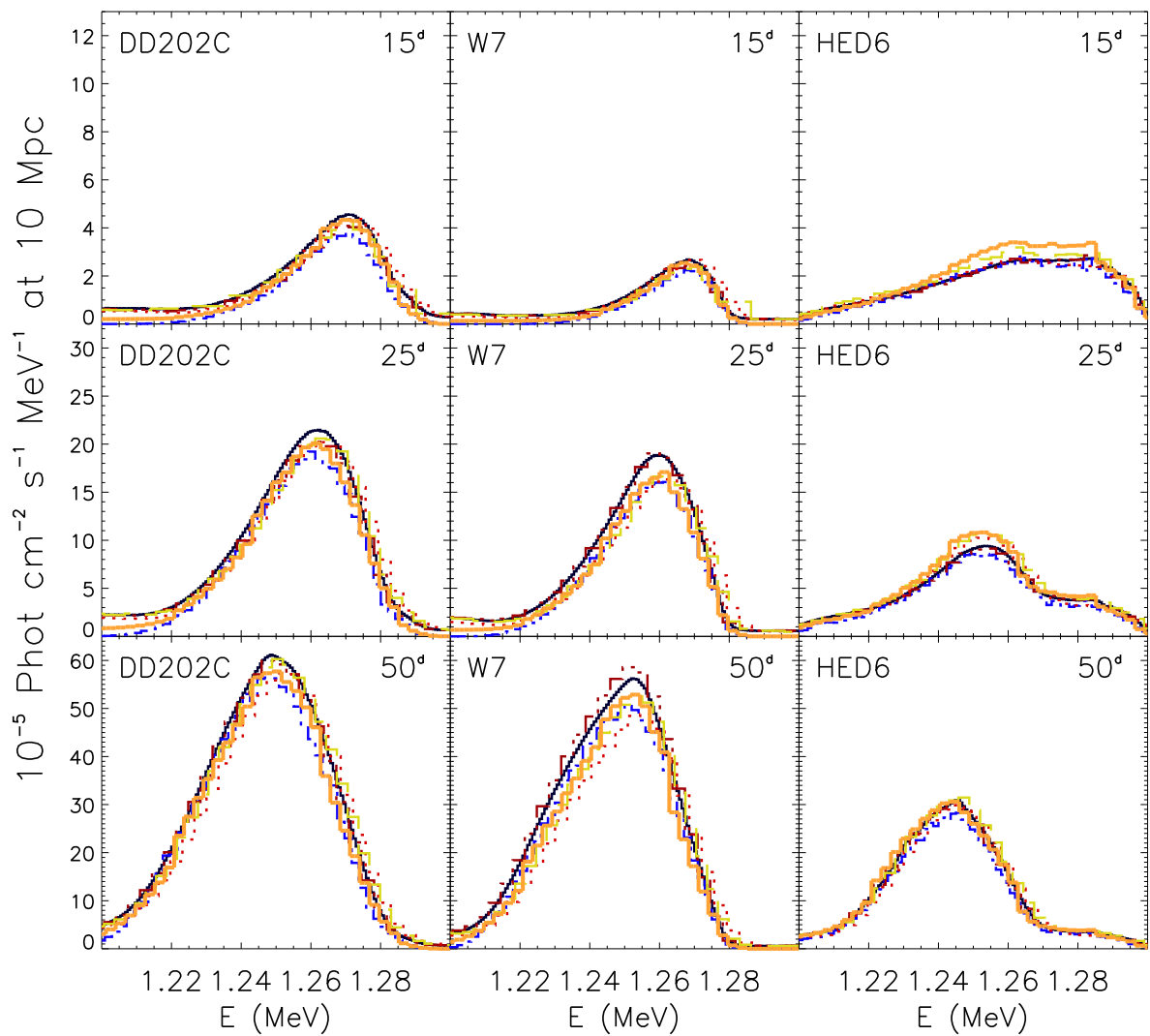}
\caption{Line profiles of the 1238 keV line for the SN Ia 
models, DD202C, W7, HED6. Although the simulations show 
noticable variations, the differences between the 
Chandrasekhar-mass models (DD202C \& W7) and the 
sub-Chandrasekhar-mass model (HED6) greatly exceeds 
the variations between simulations. Differentiating 
between DD202C \& W7 is more difficult, but is not 
rendered impossible by the variations between simulations 
if a sequence of spectra were available for comparison.}
\label{zm1238}
\end{figure}

\begin{figure}
\epsscale{1}
\plotone{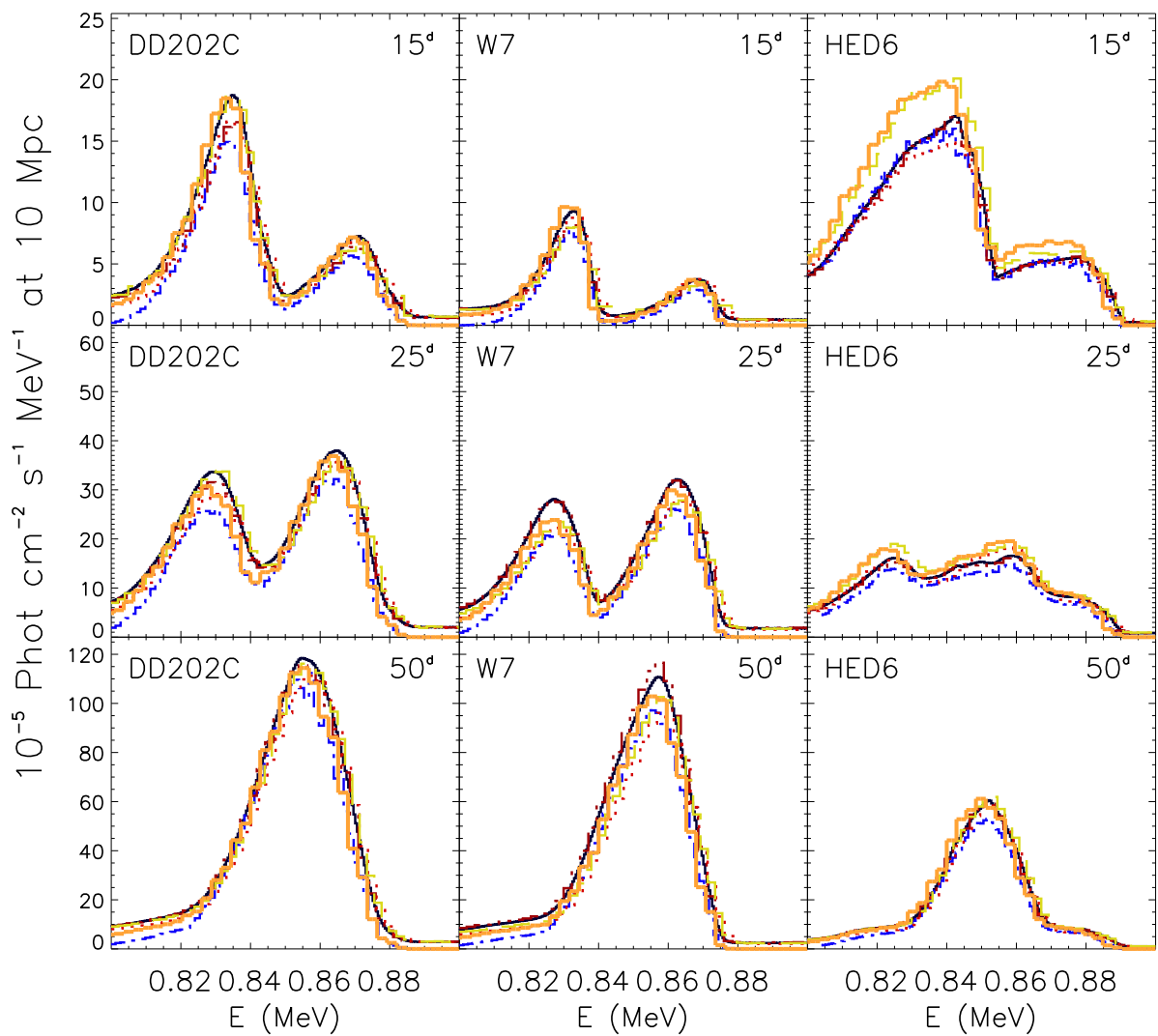}
\caption{Line profiles of the 812 \& 847 keV line complex 
for the SN Ia models, DD202C, W7, HED6. The interpretation 
is similar to that of the 1238 keV line: the differences between 
the Chandrasekhar-mass models (DD202C \& W7) and the 
sub-Chandrasekhar-mass model (HED6) greatly exceeds
the variations between simulations, and while differentiating
between DD202C \& W7 is more difficult, it is not
rendered impossible by the variations between simulations if a sequence
of spectra were available for comparison. }
\label{zm812_847}
\end{figure}

In Figures \ref{zm1238} and \ref{zm812_847} we show line profiles 
of the 1238 keV line and the 812 \& 847 keV line complex.  
The Boggs simulations are specifically of line profiles, 
and thus they contribute only to these two figures and not the 
previous three. The Kumagai and The codes did not produce 
line profiles and are thus not included in these figures. We note 
that Burrows \& The (1990) did simulate line profiles by adopting a 
technique explained in Bussard et al. (1989), which is similar to the 
technique explained in Chan \& Lingenfelter (1987). 

The Boggs line profiles, shown in Figures \ref{zm1238} and
\ref{zm812_847}, do not include the Compton scattered photons from
higher energy nuclear lines. The fact that the Boggs line profiles
agree very well with the other line profiles suggests that treating
the Compton downscattered photons has only a small effect on the line
profiles. These photons would only become important if an instrument's
energy resolution is poor enough that it samples beyond the energy
ranges shown in these figures.  

Although detailed line profile observations require instrument
sensitivities beyond those currently available (for all but
the nearest supernovae), their diagnostic potential for distinguishing
between Ia explosion models is very strong.  Because the line photons
arise primarily from non-interacting gamma-rays, the line shape is a
direct probe of the spatial distribution of $^{56}$Ni synthesized in
the supernova explosion.  For a more detailed discussion of the potential
for such observations with current and planned missions, see HWK98.

\subsection{Line Fluxes} 

A far easier observation to make, and the quantity more frequently
published from theoretical simulations, is the time evolution of
integrated line fluxes (gamma-ray light curves).  Since the Kumagai
and The codes do not include ejecta velocity effects, they compare
line emission with the other codes only through integrated flux
values, obtained by tallying ``tagged'' line photons (i.e. a photon
created at the gamma-line energy is tagged as such and contributes to
the integrated flux if it escapes with no interaction).  

Such comparisons of the lightcurves from previously published results
in HWK98 (for DD202c and HED6) revealed significant differences in
the magnitude and shape of the 812, 847 and 1238~keV lightcurves from
the results presented here.  Further inspection of the overall spectra
from HWK98 confirmed that the spectra were similar in shape, but
tended to be brighter by an epoch-dependent factor.  Closer study of
the H\"oflich code determined that a post-process step, required for
correct weight normalization of the Monte Carlo packets, was performed
incorrectly in the HWK98 spectra.  (For details, see erratum for
H\"oflich, Wheeler \& Khokhlov 1998, in press.)  When corrected for
the appropriate weight factor, which was equal to the total escape
fraction for each epoch, the HWK98 spectra roughly agree with the
spectral results in this work.

Lightcurve results from Kumagai \& Nomoto (1997) for model W7 
also demonstrated
an enhanced flux level, although the lightcurve shape was similar to
the results found here.  Comparisons with previously published W7
spectra (Kumagai \& Nomoto 1997; Kumagai, Iwabuchi \& Nomoto 1999;
Iwabuchi \& Kumagai 2001) reveal consistent results with the overall
spectra presented in $\S$3.1.  This points to an offset problem in the
generation of the integrated flux data, possibly related to setting
the SN at a given distance and/or scalings in the $^{56}$Ni mass of
the explosion model.

\subsubsection{1238 keV Line Flux}

\begin{figure}
\epsscale{0.52}
\plotone{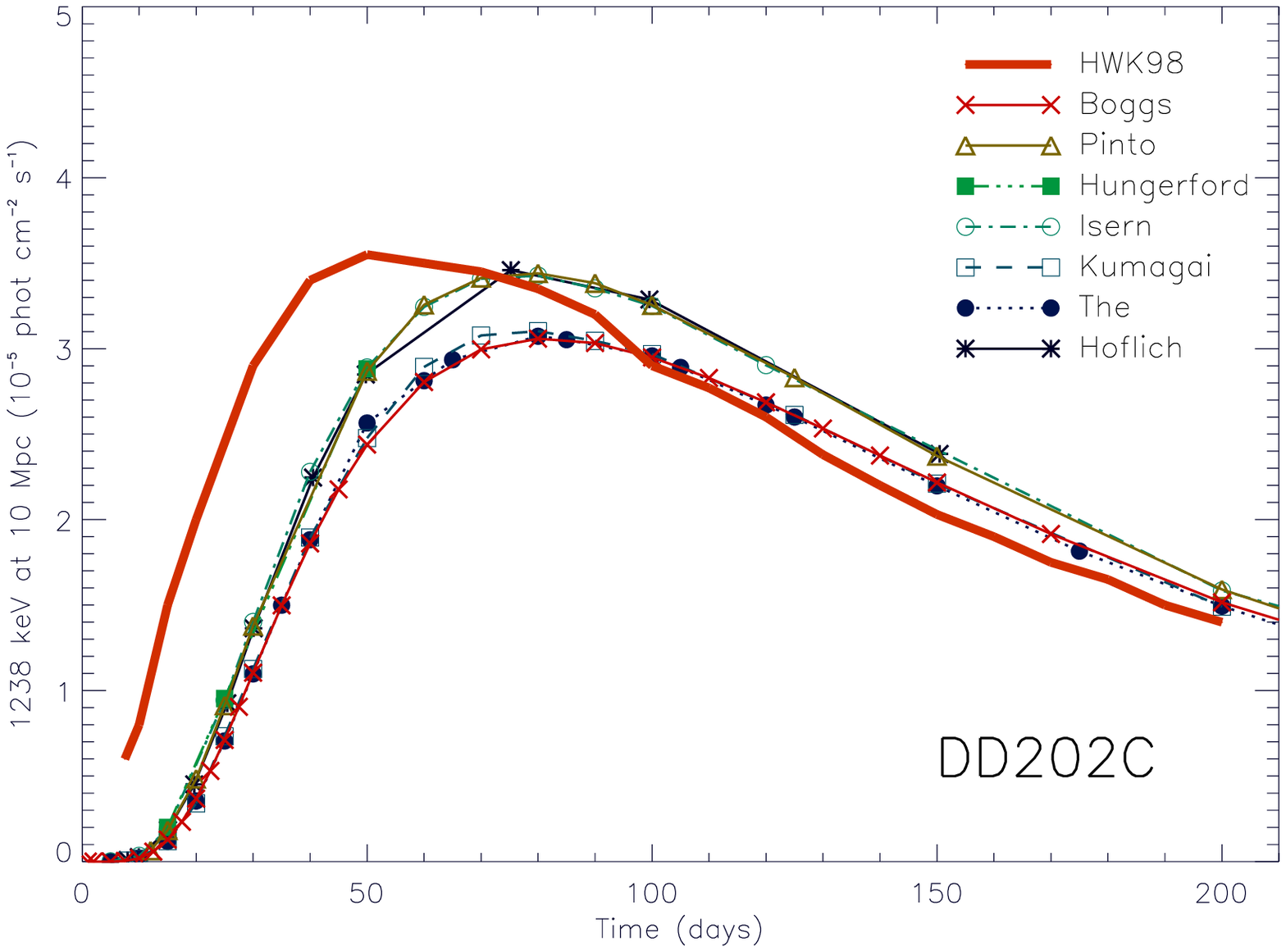}
\plotone{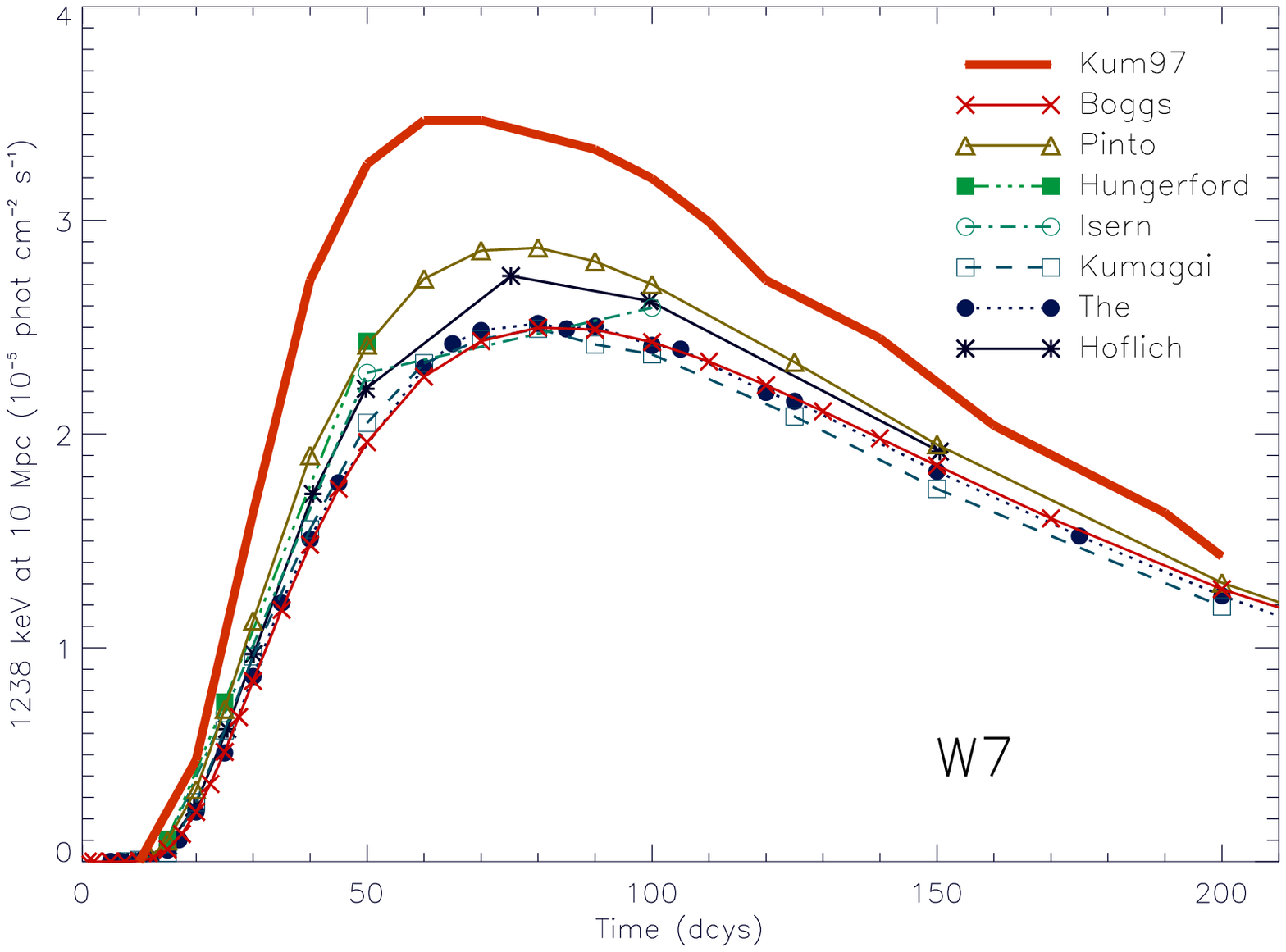}
\plotone{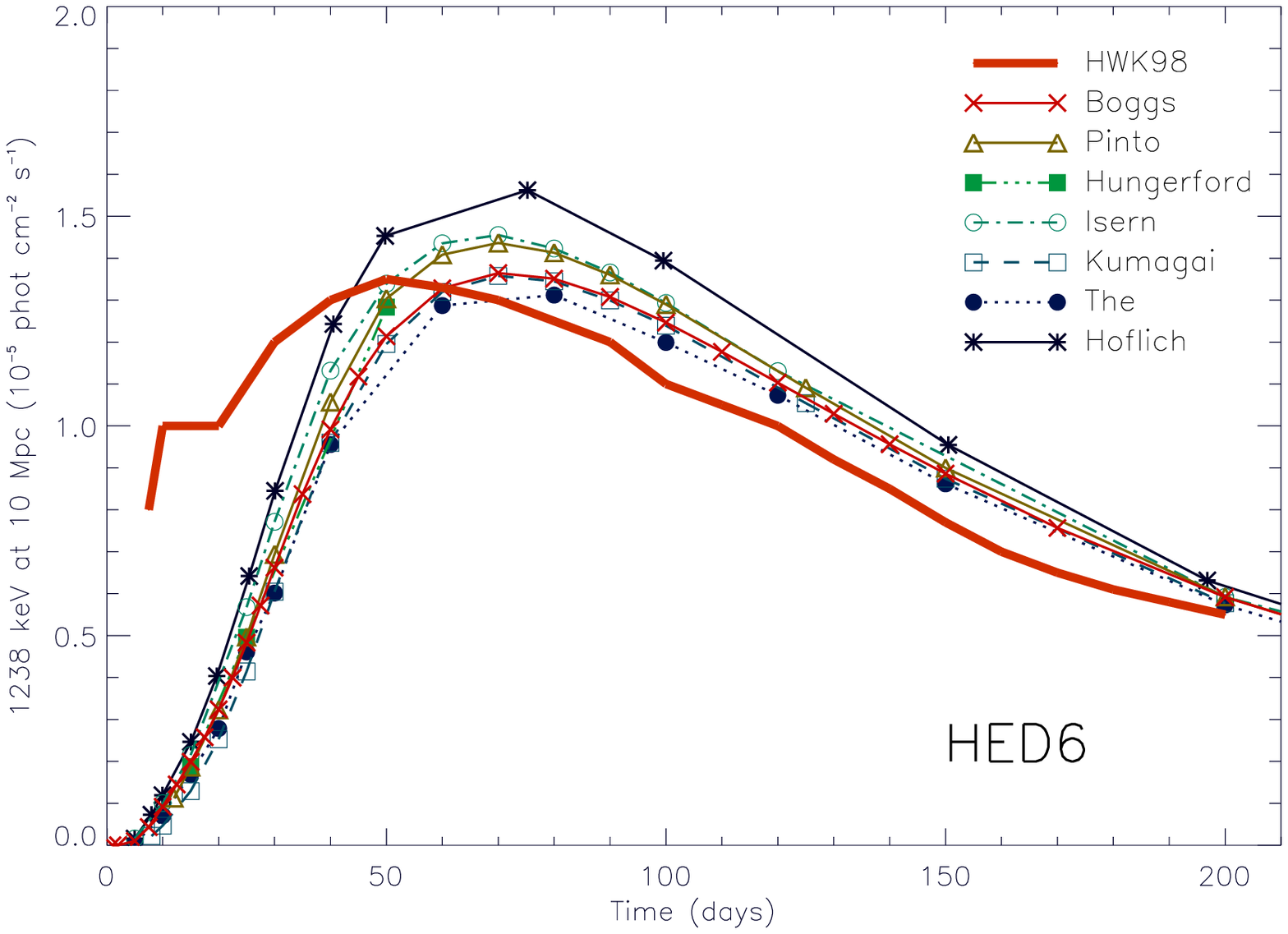}
\caption{Line fluxes of the 1238 keV line for the 
SN models, DD202C (upper panel), W7 (middle panel) and 
HED6 (lower panel). The line fluxes extracted from 
the spectra (H\"{o}flich, Maverick, Fastgam, Isern) 
agree with the line fluxes that result from tagging 
line photons (The, Boggs, Kumagai). All current 
simulations predict fainter light curves than shown in 
previous published results (HWK98 for DD202C \& HED6; 
Kumagai 1997 for W7). Spectral extraction 
assumed a 1150~-~1300 keV bandwidth. The HWK98 results are 
shown with and without the scaling for the escape fraction 
and branching ratios. 
Although the line definition in HWK98 differs from that 
used in this work, the light curves are similar when the 
corrections are applied.}
\label{f1238}
\end{figure}

The 1238~keV $^{56}$Co decay line is the most straightforward
line flux to study.  This line is isolated from other lines and there
is little continuum emission to contaminate line flux estimates.  We
define the 1238 keV line to be all photons with energies between 1150
- 1300 keV. Shown in Figure \ref{f1238} are the 1238 keV light curves
for DD202C, W7 and HED6. For comparison, we include earlier light
curves from HWK98 and Kumagai \& Nomoto (1997), 
although those works did not use
the same line definitions used in this work.

The HWK98 light curves (DD202c and HED6) are enhanced at early times
and slightly fainter than the current simulations at late times,
demonstrating the trends from the missing weight normalization
(discussed above) and the lowered $^{56}$Co decay branching ratios
(see $\S$2.2.1).  The Kumagai \& Nomoto (1997) light curve for W7 
appears too 
bright at all epochs, consistent with some offset injected during the
calculation of integrated line fluxes.

The three codes that derive line fluxes from tagged photons (The,
Kumagai \& Boggs) yielded similar light curves to the other four codes,
which obtained line fluxes from spectral extraction techniques.  This
suggests that the extraction of the line flux from the spectra can 
be performed in a manner that does
not introduce appreciable systematic errors in the light curves. It is
worth reiterating that ultimately spectra must be compared with
observations in order to infer the nickel production from an actual
supernova, so the fact that the line fluxes were adequately extracted
from the spectra is encouraging for the astrophysical use of these
simulations.

\subsubsection{812 keV and 847 keV Line Fluxes}

The two brightest gamma-ray lines occur at 812 keV and 847 keV. 
The former is produced by $^{56}$Ni~$\rightarrow$~$^{56}$Co decays, 
while the latter is produced by $^{56}$Co~$\rightarrow$~$^{56}$Fe decays.
The high-velocity expansion of the ejecta creates Doppler 
broadening that blends the two lines. Ultimately, when observed with 
an instrument that can resolve the spectra, these line profiles will provide 
a wonderful diagnostic of the nickel distribution. However, the line 
blending makes quantitative line flux comparisons between codes more 
difficult. Rather than try to isolate the individual 
contributions from each line based on the line profile, we have chosen to 
combine the two lines. Explicitly, we have defined the total flux to be 
all photons with energies between 810 - 885 keV (ignoring the fact that we 
include contamination from continuum emission). We assume equal escape 
fractions (a reasonable assumption for two lines very near in energy), 
and assign the individual line fluxes by the relative decay rates for 
each line (which are known at each epoch). For example, at 20 days the 
decay rate of $^{56}$Ni~$\rightarrow$~$^{56}$Co decays is 1.83 times the 
decay rate of $^{56}$Co~$\rightarrow$~$^{56}$Fe decays. Thus, we assign 
65\% of the total flux to the 812 keV line and 35\% to the 847 keV line. 

\begin{figure}
\epsscale{0.52}
\plotone{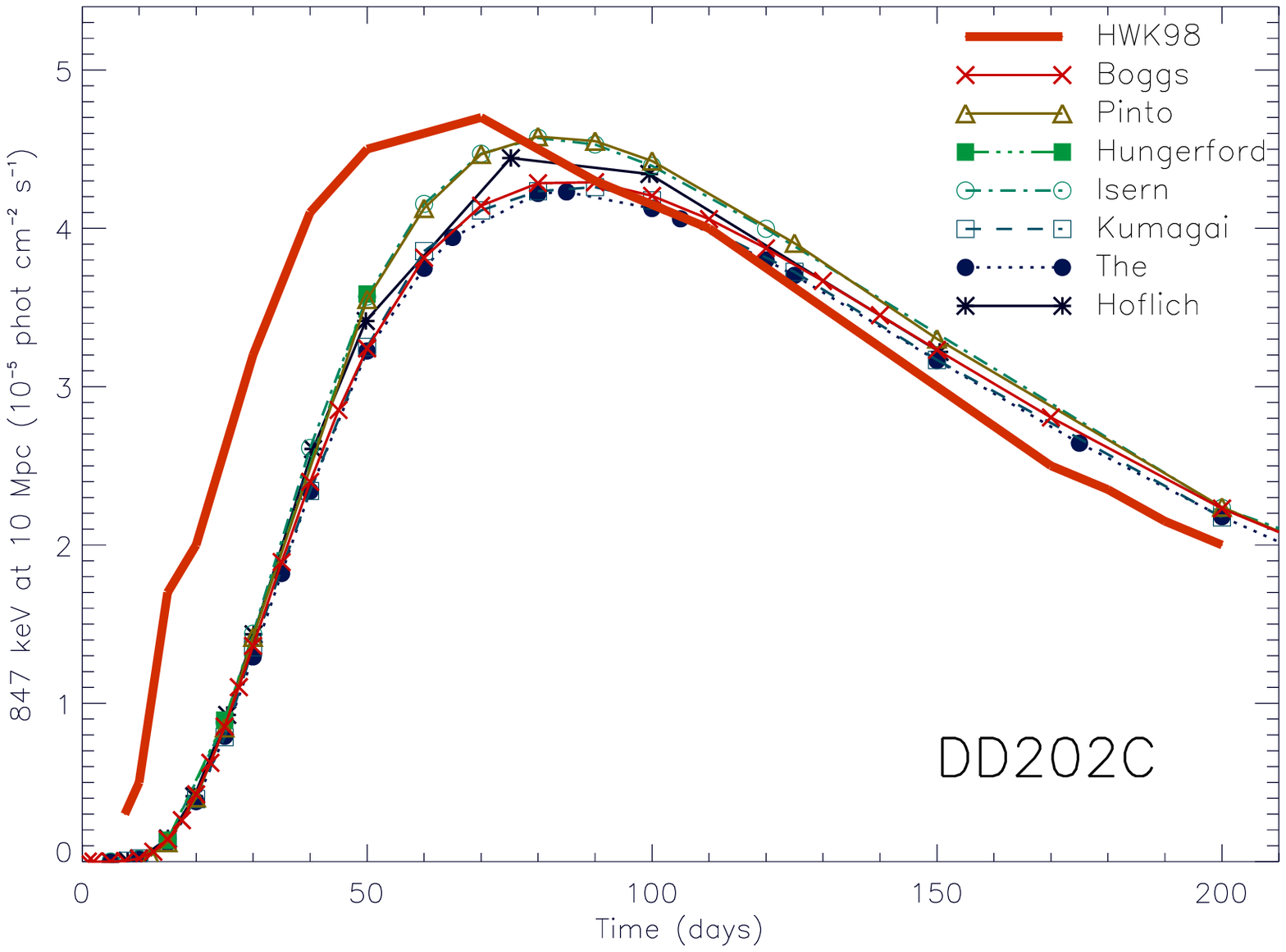}
\plotone{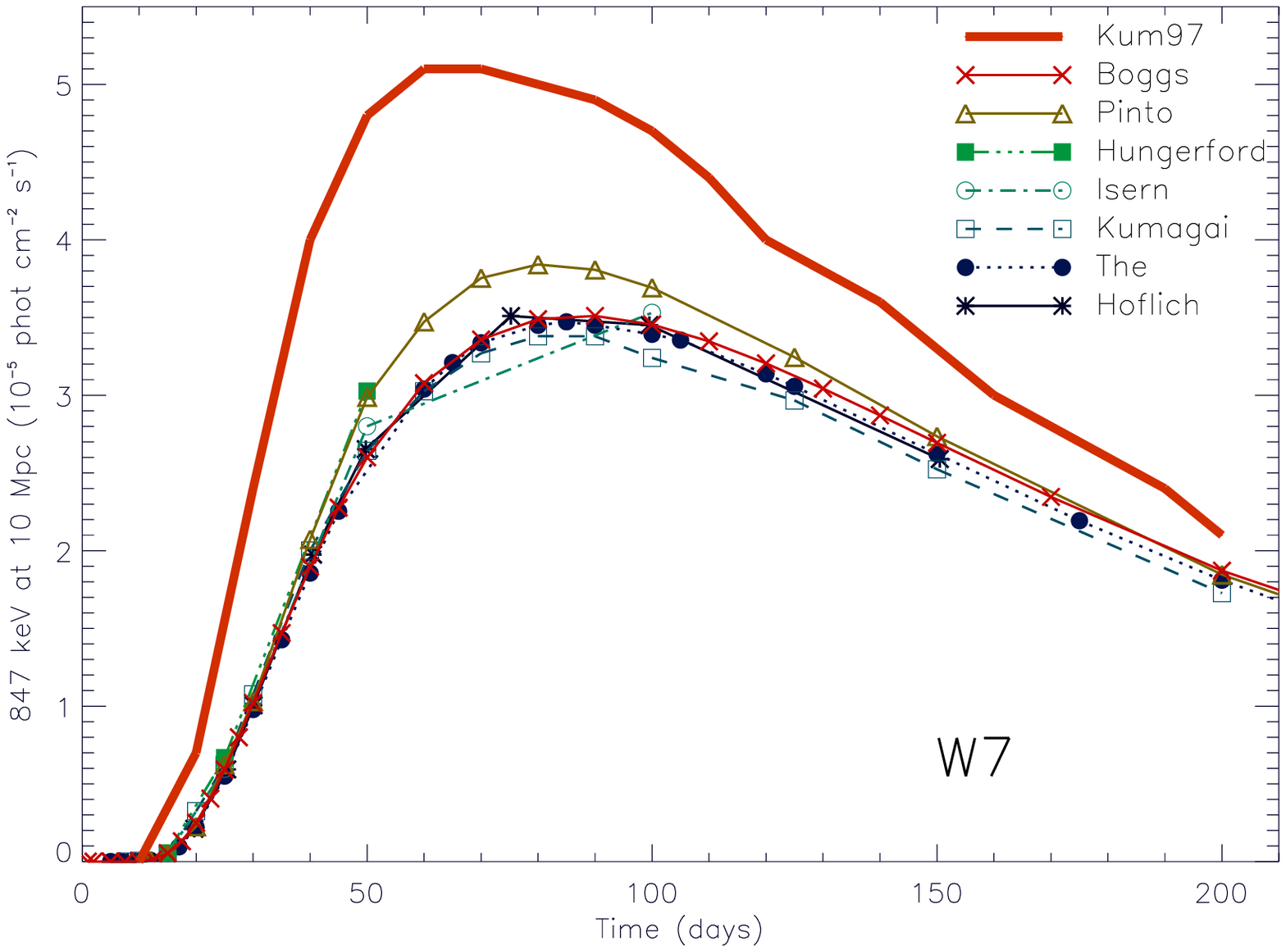}
\plotone{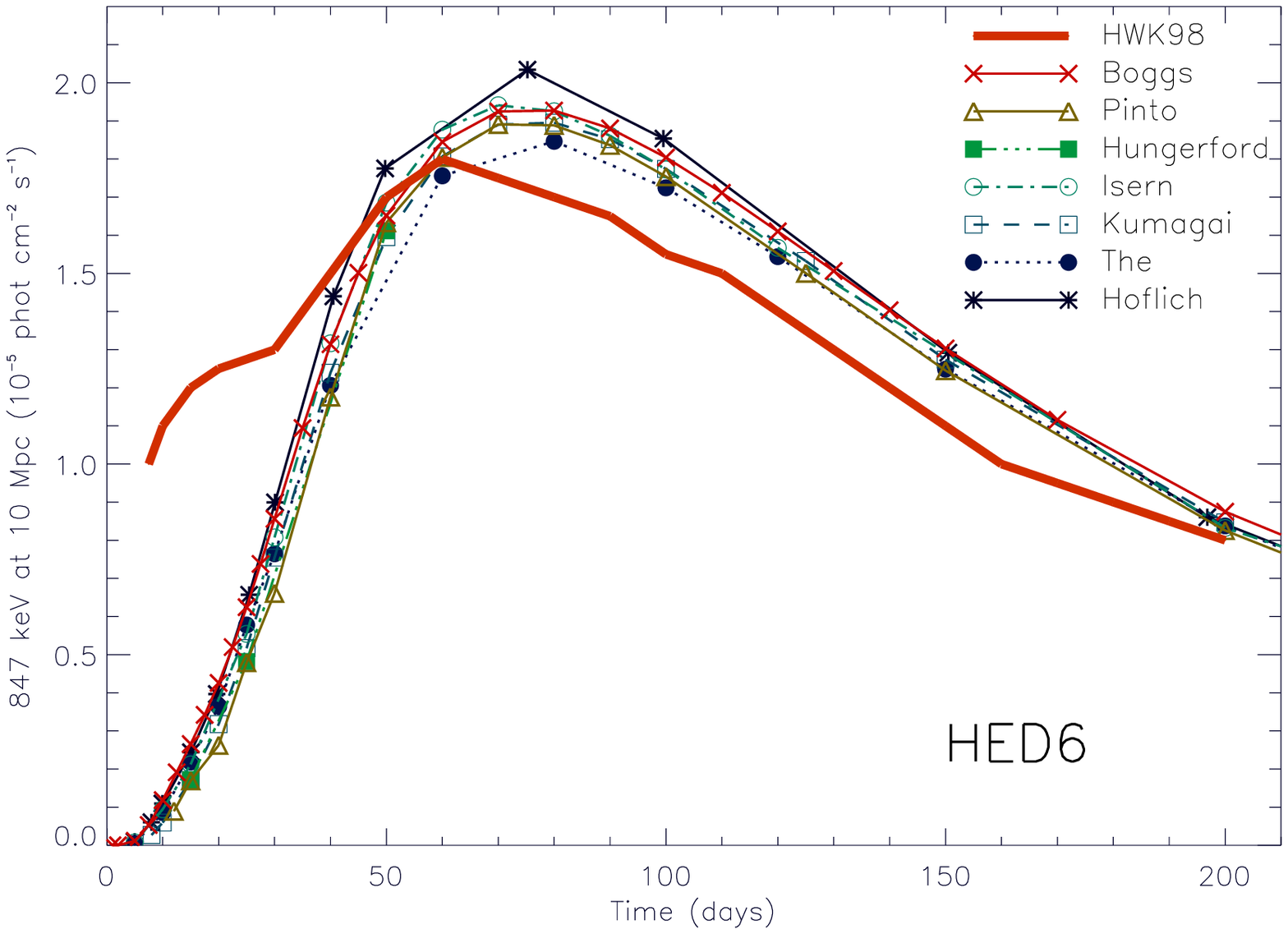}
\caption{Line fluxes of the 847 keV line for the
SN models, DD202C (upper panel), W7 (middle panel) and 
HED6 (lower panel). Spectral extraction was more complicated 
for the 847 keV line than for the 1238 keV line (requiring 
the assumption that the 847 \& 812 escape fractions are 
equal, and that all emission in the 790~-~900 keV energy 
band is line emission), but the light curves agree well with 
the light curves that result from tagging photons. 
Again, all current simulations suggest less line emission than 
suggested in HWK98 \& Kumagai 1997. Also, the scaling for 
escape fraction and branching ratios brings the HWK98 light 
curves into rough agreement with the other light curves.}
\label{f847}
\end{figure}

\begin{figure}
\epsscale{0.6}
\centerline{\plotone{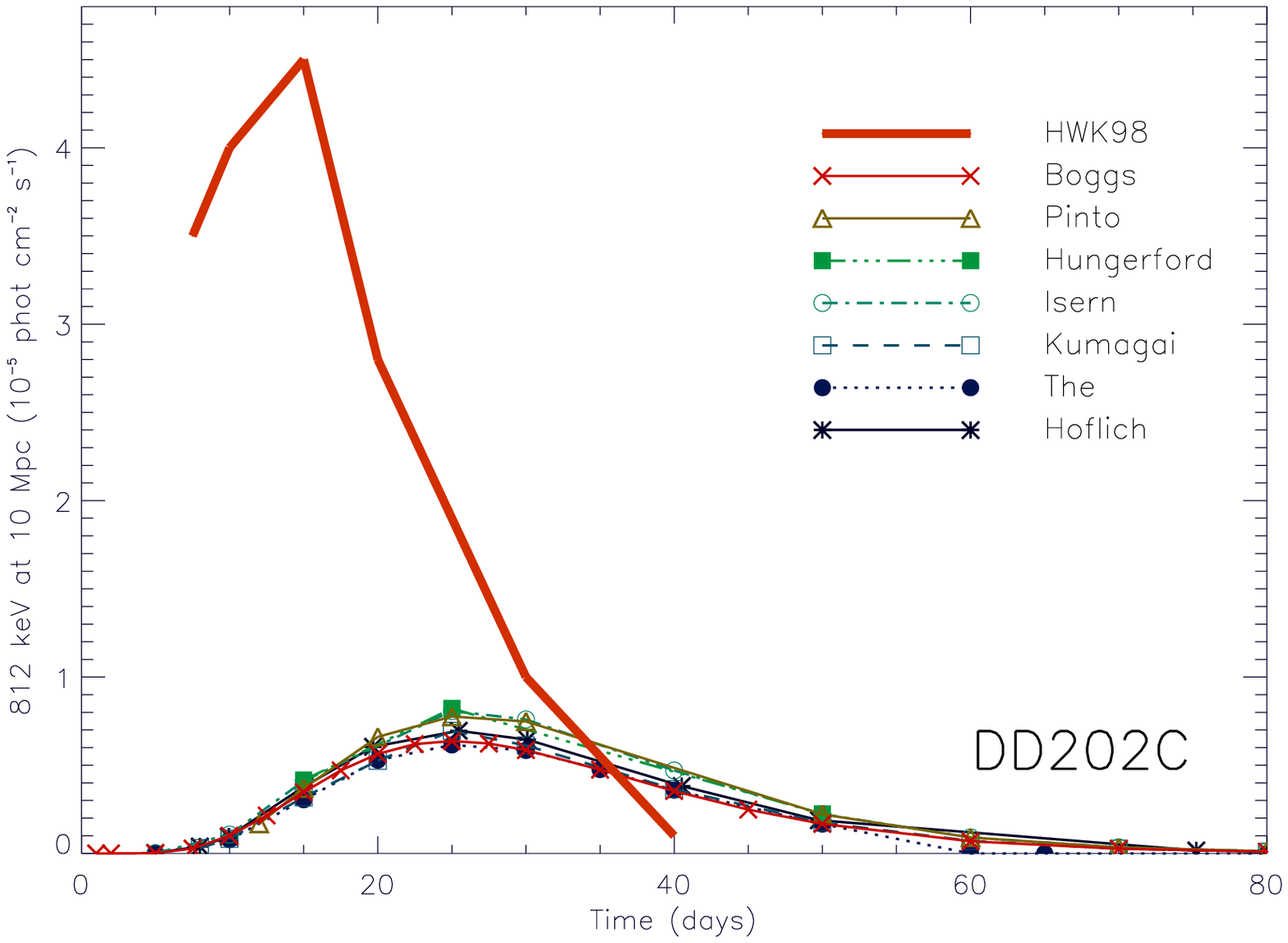}}
\centerline{\plotone{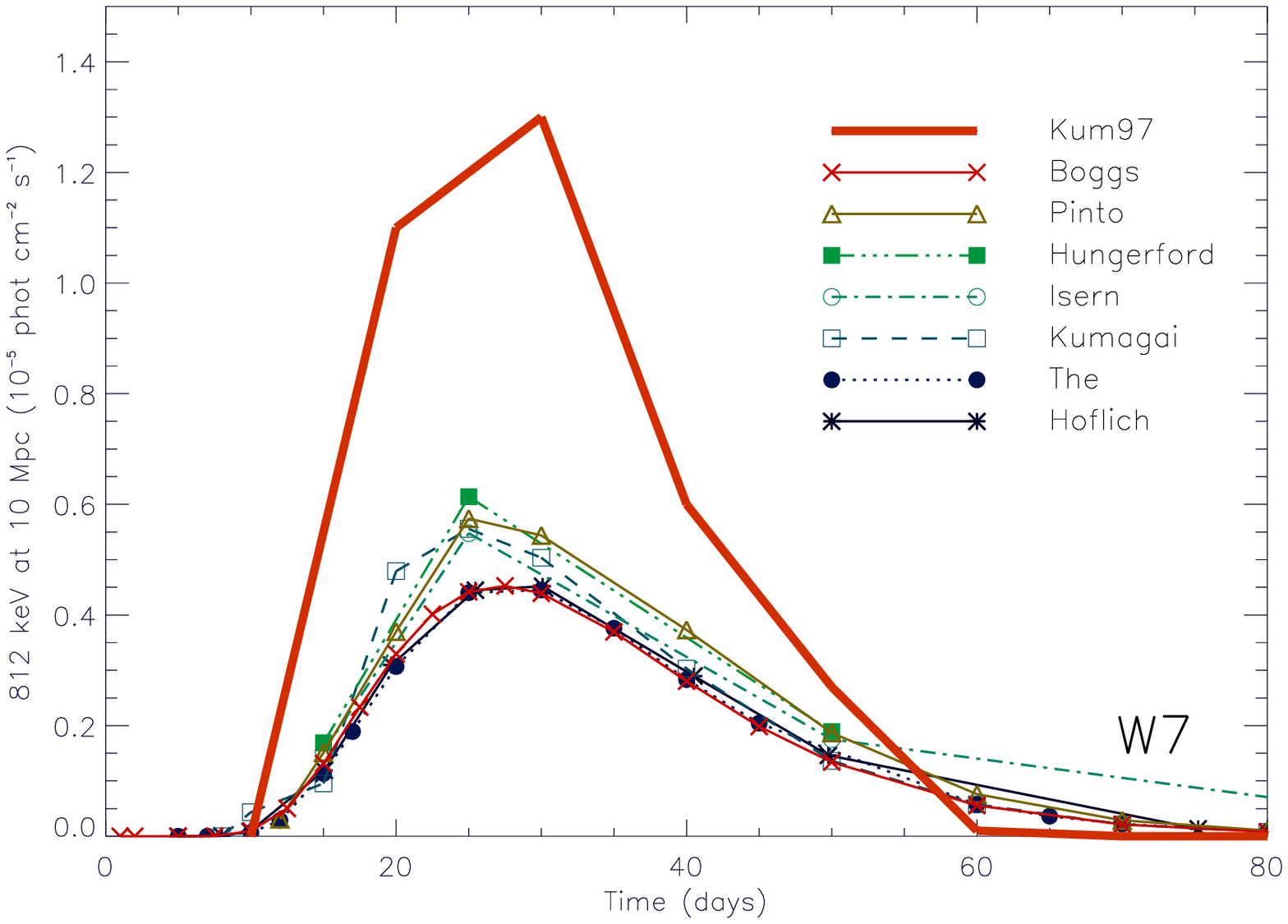}}
\centerline{\plotone{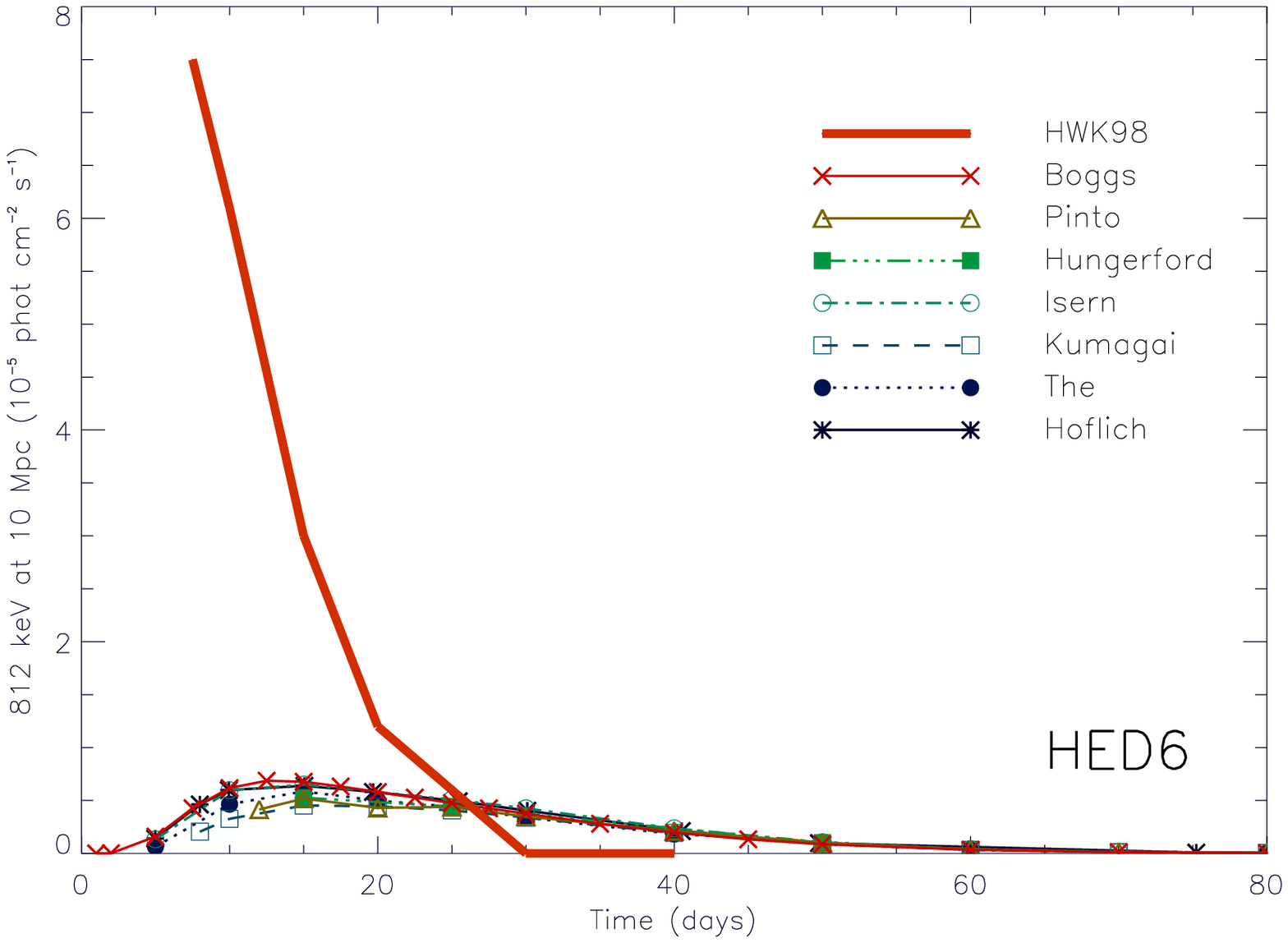}}
\caption{Line fluxes of the 812 keV line for the
SN models, DD202C (upper panel), W7 (middle panel) and
HED6 (lower panel). As with the 812 keV line emission, 
the spectral extraction and tagging light curves agree, 
and are fainter than the HWK98 \& Kumagai 1997 light curves. 
With the scaling for escape fraction and branching ratios, 
the HWK98 light curves agree fairly well with the other 
light curves. The HWK98 light curves after 20-30 days fall to zero, 
faster than the other light curves; this is due to the different 
definition for the 812 keV line employed in that work.} 
\label{f812}
\end{figure}

In Figures \ref{f847} and \ref{f812}, we show the 847 keV and 812 keV
line fluxes for the three models as simulated by all seven
codes. Again for comparison, we include earlier light curves from
HWK98 and Kumagai \& Nomoto (1997).  
The deviation at late times ($>25$~days) for
the HWK98 812~keV light curve is consistent with the lower adopted
branching ratio used in that code (see $\S$2.2.1).  As with the 1238
keV light curves, we find the same good agreement between the current
code results.

\subsection{Summary of Comparisons}

In light of the previous differences in simulated SN~Ia gamma-ray
spectra, the agreement demonstrated in this comparison is strongly
encouraging.  The differences between the individual simulations are
generally at the 10-20\% level, much less than the differences that
result from a range of input explosion models.  This is particularly
apparent in the nine panels of Figures \ref{zm1238} and
\ref{zm812_847}.  There would be no ambiguity as to which is the
correct scenario if these three models were compared with actual
observations of sufficient sensitivity.  While it is true that very
similar models might be unresolvable due to the current variations
between simulations, the level of accuracy required to perform this
type of observation will not be realized in the forseeable future.

Since we have chosen a set of explosion models that probably 
represent the full range of SNe Ia explosions, these models provide an
ideal testing ground for gamma-ray transport codes and it is likely
that codes that get good agreement against the spectra and light curves
presented here can be trusted using different explosion models as well.

Having demonstrated that the simulations have converged upon similar
solutions for these three models, we explore the range of SN Ia events
considered possible ($\S$4) and compare these simulations with
observations ($\S$5).

\section{SN Ia Line Fluxes}

With the current agreement of all seven codes for a range of explosion
models, we can now use the simulated gamma-ray signal to predict
observational differences between the explosion models.  Over the next
few years, the challenge in gamma-ray observations will be to make a
detection of a single, time-averaged flux (requiring a lengthy 
exposure).  The dominant, 847 keV
line flux peaks 50 or more days after the SN explosion, so there is ample
time for the SN to be detected and identified through optical
observations before gamma-ray observations must commence.  The 812 keV
line evolves on a shorter timescale (10-35d) and has a fainter peak
(limiting its detection to very local supernovae).  As the SN takes
roughly the same timescale to reach the optical peak, gamma-ray
observations need to commence a few days before optical peak to
contain the 812 keV peak. A large fraction of nearby SNe Ia are
detected at peak or later, so this requirement places strict demands
upon ``target-of-opportunity" telescopes.  
 
In this section we show line flux light curves for a collection
of SN Ia models simulated with the The code.  We separate the
models into three sub-classes based on observational categories:
normally-luminous, sub-luminous and super-luminous\footnote{ 
Although we do not use this information, we mention that Li et 
al. (2000) assert that roughly 60\% of SNe Ia are considered 
normally-luminous, 20\% sub-luminous and 20\% super-luminous.}.

\subsection{Normally-luminous SNe Ia}

\begin{figure}
\epsscale{0.6}
%\centerline{\plotone{ps/norm.ps}}
%\centerline{\plotone{ps/super.ps}}
%\centerline{\plotone{ps/sub.ps}}
\centerline{\plotone{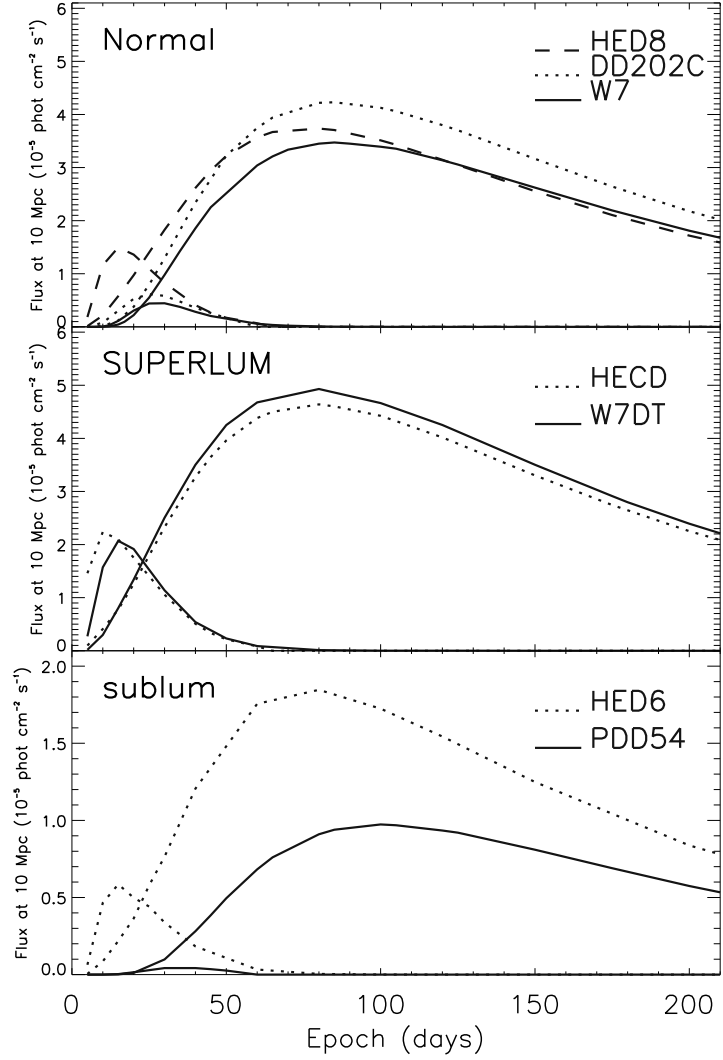}}
\caption{Line fluxes of the 812 \& 847 keV lines for 
SN models representative of the 3 luminosity sub-classes. 
The super-luminous models (W7DT \& HECD) have the brightest 
gamma-ray lines, but are the most homogeneous, while the 
sub-luminous models (PDD54 \& HED6) are faint but differ 
appreciably.  The Chandrasekhar-mass normally-luminous models 
(W7 \& DD202C) differ at late times due to their different 
nickel production, and the sub-Chandrasekhar mass model (HED8), 
differs early due to nickel produced very near the surface.}
\label{subclass}
\end{figure}

This is the most frequent SN Ia sub-class and the best studied. 
SN 1998bu was considered normally-luminous and is grouped in this 
category (\S 5.3).  We compare 
three models that fit within this sub-class, W7 (a Chandrasekhar-mass 
deflagration); DD202C (a Chandrasekhar-mass delayed detonation) and  HED8 
(a sub-Chandrasekhar mass helium detonation). 
The light curves are shown in the upper panel of Figure \ref{subclass}. HED8 
creates the least amount of nickel, but has nickel near the surface. 
This leads to HED8 being the brightest model of the three 
at early epochs, but the 
faintest model after 150 days. For a sufficiently early observation of a 
nearby supernova, DD202C and W7 are easily distinguished 
from HED8 based upon the 812 keV line (or equivalently, the timing of the rise 
of the 847 keV line).

\subsection{Super-luminous SNe Ia}

This SN Ia sub-class differs from the normally-luminous SNe Ia in that
the explosion creates more nickel for each scenario.  SN 1991T was
considered super-luminous and is grouped in this sub-class (\S 5.2).
We compare two models, W7DT (a Chandrasekhar-mass late detonation that
is very similar to W7 but includes additional nickel production nearer
the surface) and HECD (a sub-Chandrasekhar mass helium detonation that
is more massive and produces more nickel that HED8).  These models
produce brighter light curves (middle panel of Figure \ref{subclass}),
but the two super-luminous explosion models do not differ
dramatically, and it will be difficult to distinguish them based on
the gamma-ray light curves alone. The result is that this type of
explosion is detectable to large distances, but is not distinguishable
to a comparatively large distance.

The super-luminous models are characterized by nickel near the surface
of the ejecta.  While this leads to 812 keV emission at earlier epochs
than predicted for normally-luminous SN Ia models, the 812 keV peak is
much lower than suggested in HWK98 and Kumagai \& Nomoto (1997). The 
largest deviations between past works and this current work occur in this
``super-luminous'' type Ia subclass.

\subsection{Sub-luminous SNe Ia}

This sub-class is the least promising for gamma-ray studies.  SN 1986G
was considered a slightly sub-luminous event and is best (though
imperfectly) grouped in this sub-class (\S 5.1).  Sub-luminous events
are less frequent than normally-luminous SNe Ia and produce much
fainter gamma-ray emission.  For Chandrasekhar-mass explosions, the
nickel production is very low and is all concentrated near the center
of the supernova. This results in extremely faint gamma-ray emission.
Sub-Chandrasekhar mass explosions also produce very little nickel, but
occur in lower mass objects, so the high escape fractions partially
compensate for the lower nickel production.  We compare two models,
PDD54 (a Chandrasekhar-mass pulsed delayed detonation) and HED6 (a
very low-mass helium detonation). Different Sub-luminous models
produce quite different light curves, but all are so faint that they
will be difficult to detect (lower panel of Figure \ref{subclass}).

\section{Observed SNe Ia}

In the last 25 years, there have been three SNe Ia that were 
close enough to warrant observations with gamma-ray 
telescopes.\footnote{The sub-luminous SN Ia, SN 2003gs, was 
observed with the SPI instrument on the INTEGRAL satellite. The 
analysis of those observations has not been completed.} 
 Although none of the three resulted in significant 
detections, papers have been written that infer the nickel 
production in each SN based upon the observations. We revisit 
these three observations and discuss to what level they 
constrain the potential explosion mechanisms. 

\subsection{SMM Observations of SN 1986G}

SN 1986G was first detected in Centaurus A on May 3, 1986 (Evans 1986,
IAU Circ., No. 4280).  It was discovered one week before maximum light
and exhibited a relatively narrow luminosity peak. Its high 
$\Delta$m$_{15}$(B) value led to its classification as a 
slightly sub-luminous SN (Hamuy et al. 1996).
Heavy host galaxy extinction was suggested by both the
photometric colors and by strong Na-D absorption. Although some papers
have argued that the extinction was large enough to infer an absolute
magnitude in the normal range (Cristiani et al. 1994), recent studies
of the host galaxy extinction to SNe Ia maintain that SN 1986G was
slightly sub-luminous (Phillips et al. 1999).

The Gamma-Ray Spectrometer on-board the Solar Maximum Mission (SMM)
satellite observed the SN with sensitivity that varied from 30\% to
full sensitivity during the entire epoch of cobalt decay. Matz \&
Share (1990) derived upper limits for the 847 and 1238 keV line
emission from SMM spectra.  They used escape fractions published by
Gehrels, Leventhal \& MacCallum (1987) from a collection of
parameterized SN Ia models to derive that the upper limits for the
nickel production ranged from 0.36 - 0.41 M$_{\odot}$ (assuming a
distance of 3 Mpc to Centaurus A). This upper limit is marginally
consistent with the 0.45 $\pm$ 0.03 M$_{\odot}$ $^{56}$Ni production
(scaling the distance from 3.3 $\pm$ 0.3 Mpc to 3.0 Mpc) derived from
the nebular spectra (Ruiz-Lapuente \& Lucy 1992).

Matz \& Share quoted their results in terms of the $^{56}$Ni
production allowed by the observations.  We do not re-analyze the SMM
observations. Instead, we compare the average fluxes during the 1986
August 25~-~October 9 interval during which the SMM sensitivity was
the largest.  A review of the escape fractions from Gehrels, Leventhal
\& MacCallum (1987) confirms that their range is in agreement with the
simulations performed in this work. The SMM instrument had a
$\sim$~80~keV FWHM at these energies and thus sampled a broad range of
the continuum in addition to the two lines. However, for most of the
epochs included in the composite SMM observation, the SN would have
been expected to emit a relatively faint continuum. Thus, very little
error is introduced by using tagged line photons and ignoring the
instrument energy resolution for this SN.

\begin{figure}
\epsscale{1.2}
\centerline{\plotone{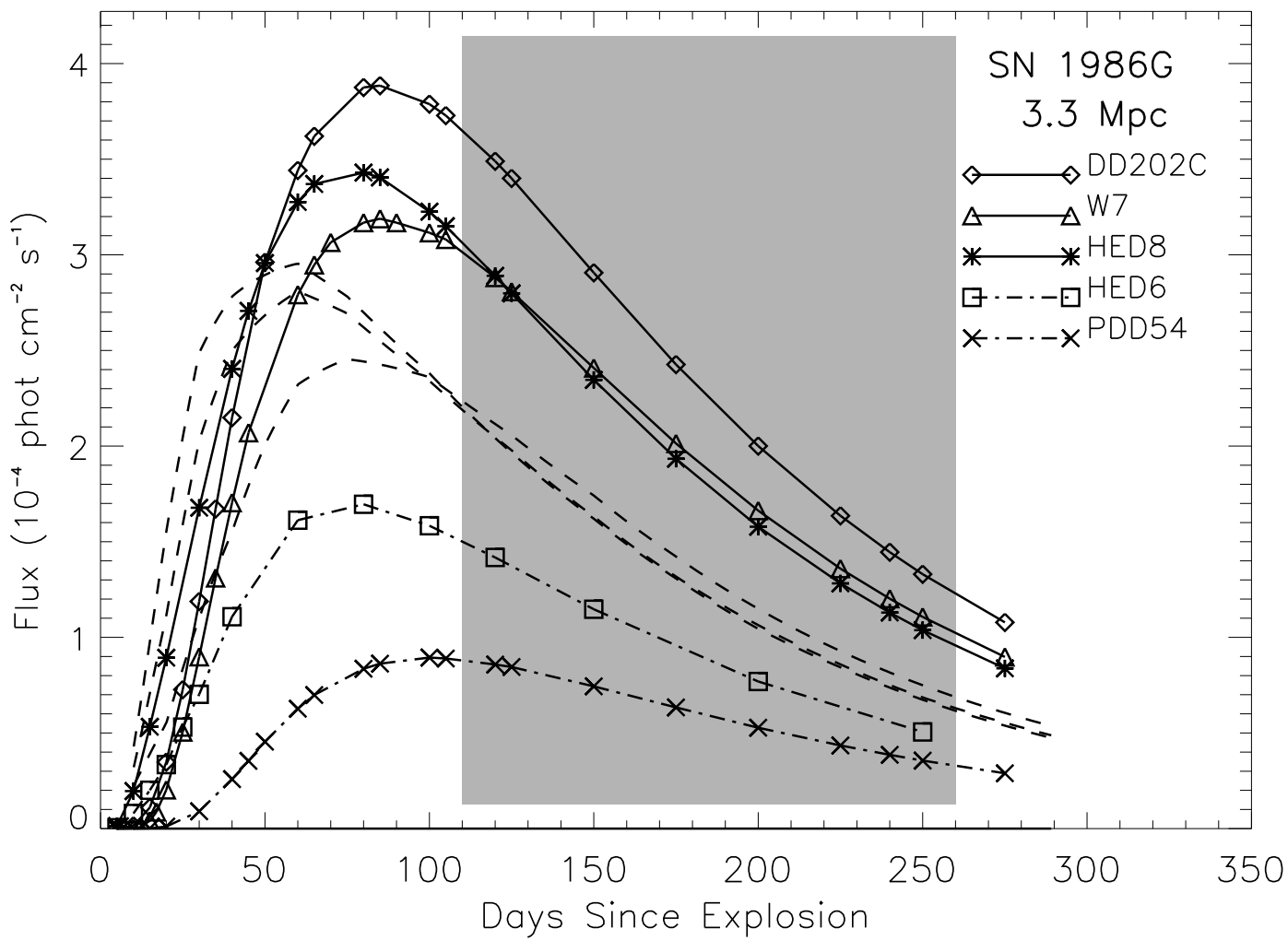}}
\caption{Five current simulations of 847 keV line emission from 
SN Ia models at 3.3 Mpc compared 
with 3$\sigma$ upper limit light curves derived from SMM observations 
of SN 1986G (Matz and Share 1988). The three Matz and Share light curves 
are shown with dashed lines, the five current simulations are identified 
as shown. The epoch of maximum SMM angular response to SN 1986G
is shaded. The normally-luminous models appear too bright at the 3$\sigma$ 
level, while the very sub-luminous models are acceptably faint.}
\label{sn1986G}
\end{figure}

In Figure \ref{sn1986G}, we compare light curves for five models with 
the light curves for the three models treated in Matz \& Share, in 
all cases setting the distance to be 3.3 Mpc. 
The three Matz \& Share light curves assume the 3$\sigma$ upper 
limit $^{56}$Ni masses, the other models use the published masses 
(as listed in Table \ref{tab_mods}). 
While only the 847 keV line emission is shown in the figure, the upper 
limit $^{56}$Ni masses were based upon a joint 847-1238 keV line fit. 
The figure shows that the three normally-luminous SN Ia models (DD202C, 
W7 and HED8) all produce too much 847 keV emission, while 
the very sub-luminous SN Ia models are faint enough to remain 
below the upper limits, especially for the low-nickel PDD54.  

Note that all of these light curves assume the distance to 
Centaurus A to be 3.3 Mpc. Measures of this distance arrive at 
3.1 $\pm$ 0.1 Mpc (Tonry \& Schecter 1990) and 3.6 $\pm$ 0.2 Mpc (Jacoby et al. 1988), 
suggesting that slightly more $^{56}$Ni production could be permissible. 
Thus, it appears that SN 1986G was tantalizingly close to being detected 
by SMM, and it would have been detected had it been a normally- or 
super-luminous event rather than slightly sub-luminous. Nonetheless, the 
upper limit is consistent with the current understanding of 
SNe Ia and the simulation of gamma-ray escape from SN models. 

\subsection{CGRO/COMPTEL and OSSE of SN 1991T}

SN 1991T was first detected in NGC 4527 on April 13, 1991 (S. Knight,
IAU Circ., No. 5239) more than a week before maximum light.  Its
pre-maximum spectra featured iron-peak elements instead of the
intermediate-mass elements of normal SNe Ia supernovae, but, after
peak, it was spectro-scopically normal.  The light curves were broad
($\Delta$m$_{15}$(B) value of 0.94), leading to the suggestion that SN
1991T was a super-luminous SN Ia and it became a template slow SN Ia
(though slower SNe exist).  SN Ia models were produced explicitly
to explain the optical observations of 1991T, we have included two of
these models in this study (W7DT and HECD).

\begin{figure}
\epsscale{0.7}
\centerline{\plotone{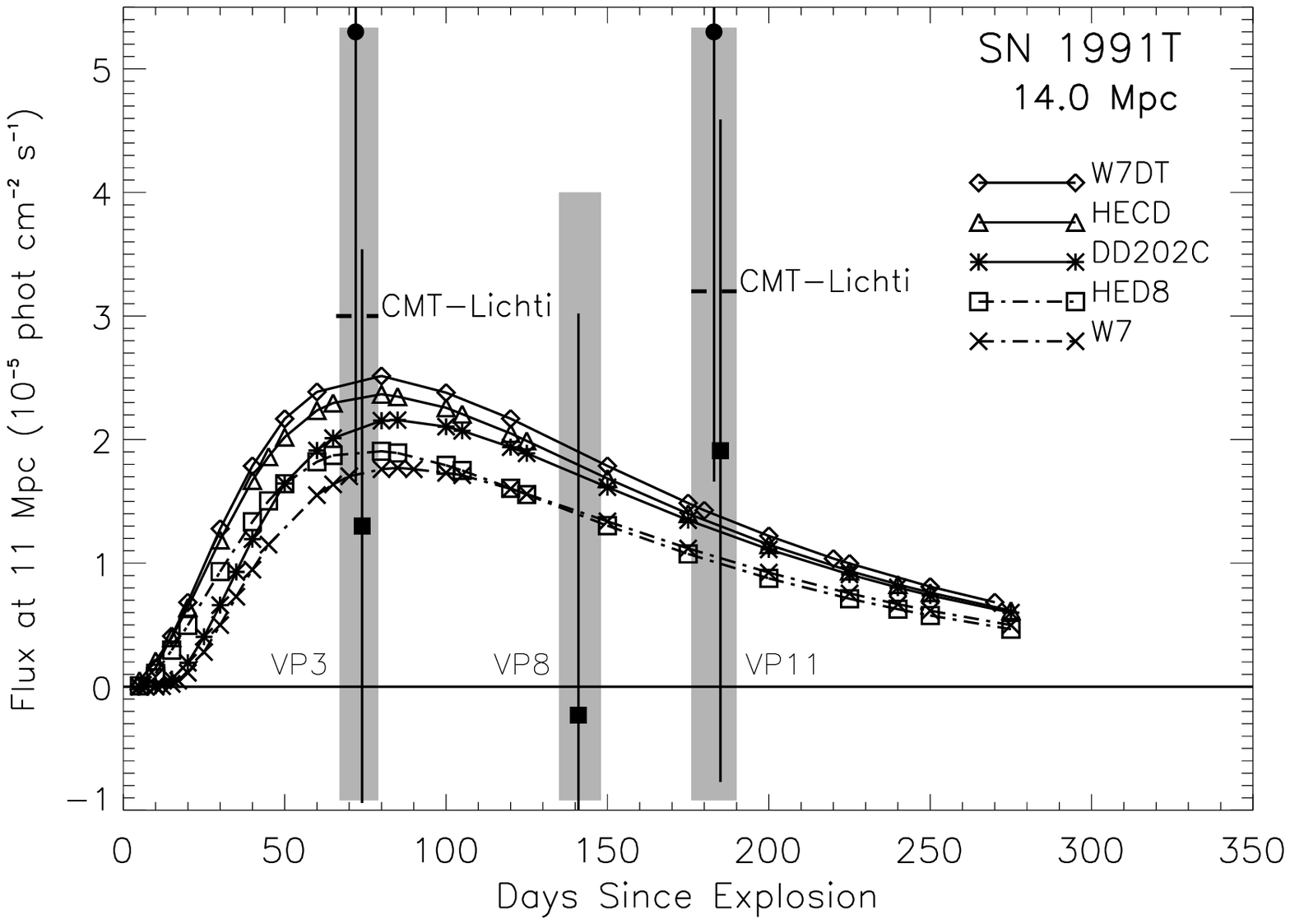}}
\centerline{\plotone{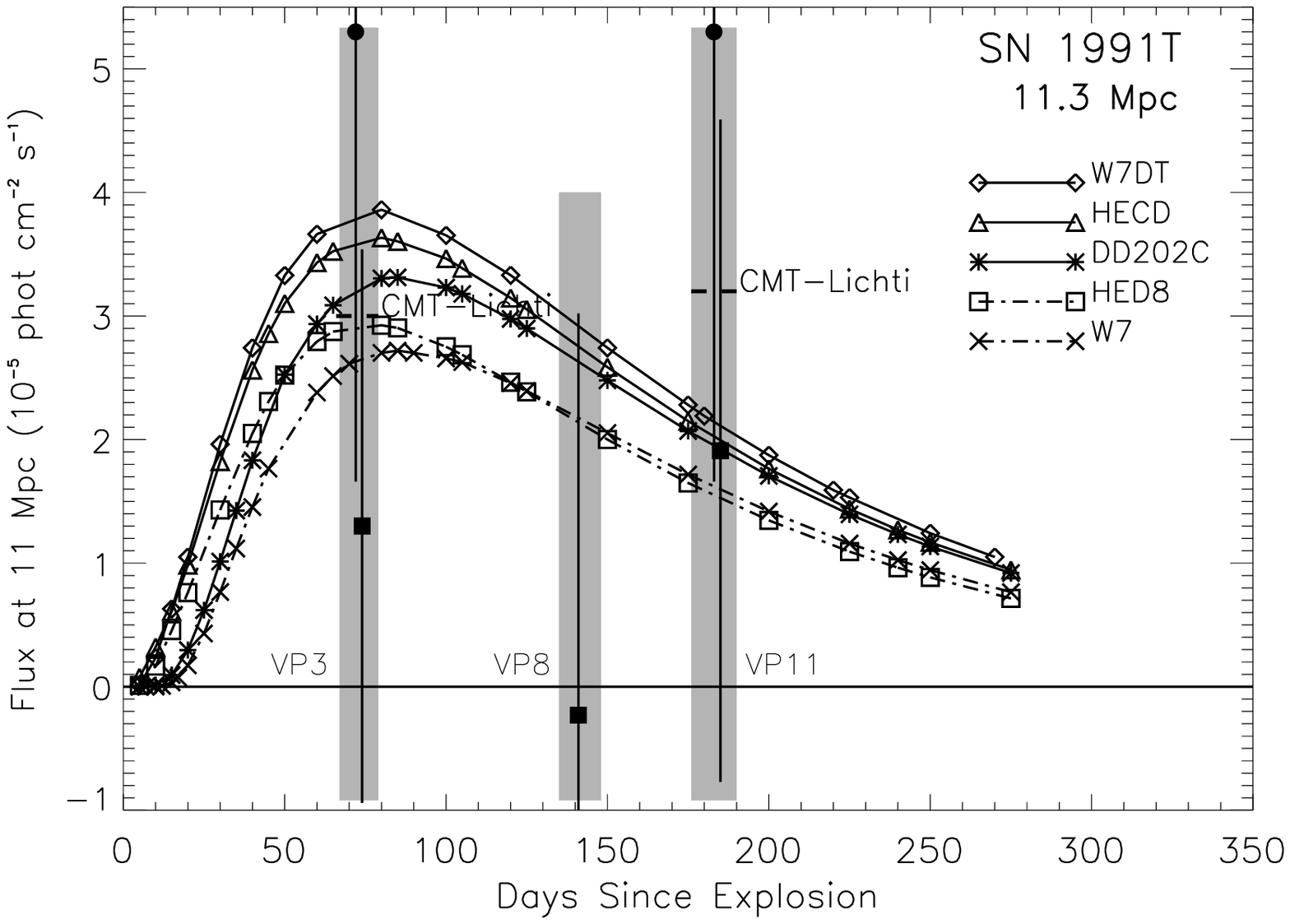}}
\caption{Five current simulations of 847 keV line emission from 
SN Ia models at two assumed distances are compared with COMPTEL 
and OSSE observations of SN 1991T. The upper panel shows the models at 
the larger distance of 14.0 Mpc, the lower panel shows the models at 
11.3 Mpc. The five current simulations are identified as shown. 
The shaded regions show the three viewing periods, VP3, VP8 (OSSE-only) 
and VP11. The OSSE data points (filled boxes) and COMPTEL-Lichti (2$\sigma$ 
upper limits, dashed lines) are fainter than the models, while the COMPTEL-Morris 
(filled circles) are brighter than the models 
(Lichti et al. 1994, Leising et al. 1995, Morris et al. 1997). 
The fluxes were all derived from joint 847/1238 keV line fits.}  
\label{sn1991T}
\end{figure}

The Compton Gamma-Ray Observatory had just been
launched (one week before the discovery of SN 1991T), and months of
calibrations and other testing had to be performed before the
instruments on-board CGRO could observe the SN. Observations were
initiated on June 15, 67 days after the explosion (assuming the SN was
detected 3 days after the explosion), and continued in three viewing
periods (3,8,11) until 190 days after the explosion (COMPTEL observed
only viewing periods 3 and 11).  There were two instruments on CGRO
that were capable of detecting the 847 and 1238 keV lines from the SN,
the COMPTEL and OSSE instruments. Separate analyses were performed
upon the two sets of observations. Initially, COMPTEL reported only
upper limits for the 847 and 1238 keV lines, arriving at 2$\sigma$
upper limits for the 847 keV line of 3.0 x 10$^{-5}$  phot $cm^{-2}$
$s^{-1}$ and 3.2 x 10$^{-5}$ phot $cm^{-2}$ $s^{-1}$ for each viewing
period (Lichti et al. 1994). A later, independent analysis suggested a
combined 3.3$\sigma$ detection (Morris et al.  1997). OSSE analysis
derived only upper limits, reporting a 3$\sigma$ upper limit of 
(4.1-6.6)~x~10$^{-5}$ phot~$cm^{-2}$ $s^{-1}$ for the 847 keV line 
during the first observation, based upon a combined, simulataneous fit 
to the 847 keV and 1238 keV lines during all three epochs 
(Leising et al. 1995). When fit separately, the formal fluxes are 
(1.3$\pm$2.2, -0.2$\pm$3.2, 1.9$\pm$2.7)~x~10$^{-5}$ phot~$cm^{-2}$~$s^{-1}$ 
for the 847 keV line for each of the three viewing periods (M.D. Leising, 
personal communication). We compare five models to the OSSE observations 
combined with each of the COMPTEL results (Figure \ref{sn1991T}).

The uncertainty of the distance to NGC 4527 has made interpretation of
the upper limits to the gamma-ray emission complicated.  Using the
distances to suggested neighbor-galaxies yielded a range of distances
from 10-17 Mpc.  With such a range, astronomers could either largely
reject the Ia models by the observed upper limits or find that almost
all models were consistent.  (see Leising et al. 1995 for an
explanation of the difficulties simultaneously explaining the optical
and gamma-ray observations of SN 1991T).  New studies have narrowed 
the distances to a range from 11.3-14.0 Mpc (Richtler et
al. 2001, Gibson \& Stetson 2001, Saha et al. 2001).\footnote{We note
that the current range of distances, combined with extinction
estimates, lead to the absolute magnitude of SN 1991T spanning the
scatter of SNe Ia about the LWR (i.e. the 11.3 Mpc distance would make
SN 1991T faint for its light curve shape, while the 14 Mpc distance
would make it slightly bright for its light curves shape).}  
For this work, we place NGC 4527 at 11.3 and 14.0 Mpc. 

OSSE did not detect emission from SN 1991T, although VP3 was very near 
the peak of the simulated cobalt line peaks. Thus, those observations 
favor models that feature low gamma-ray fluxes. However, the modest 
sensitivity of the OSSE instrument limits the ability to 
discriminate between explosion scenarios. 

The COMPTEL observations would, in principle, strengthen the ability to 
distinguish explosion scenarios. However, this is not (unambiguously) 
the case because the two separate analyses of the COMPTEL data  
arrived at dramatically different conclusions. The analysis by 
Lichti et al. (1994) detected no emission from SN 1991T, and thus favors 
models that feature low gamma-ray fluxes.  When combined with the OSSE observations, 
the COMPTEL upper limits further favor low gamma-ray flux models, 
at the level that the brighter models would be considered inconsistent 
(Leising et al. 1995). 
By constrast, the Morris et al. (1997) analysis measures 
fluxes brighter than predicted by any of the models. Using those 
fluxes, the highest flux models are favored, the more sensitive COMPTEL 
observations counteracting the OSSE upper limits. 

The inability to reconcile these datasets severely limits the physics that 
can be derived from the observations (at least at the current level of understanding 
of SN Ia explosion physics). The OSSE observations do not reject any 
of the explosion scenarios if the larger NGC 4527 distance is used, 
and the COMPTEL observations are ambiguous. 

\subsection{CGRO/COMPTEL Observations of SN 1998bu}

SN 1998bu was discovered by M. Villi on May 9, 1998 (IAU Circ.,
No. 6899) in M96 (NGC 3368) more than a week before maximum light,
affording CGRO a second opportunity to observe a SN Ia.  This SN was
determined to be a normally-luminous SN ( $\Delta$m$_{15}$(B) = 1.02
$\pm$ 0.04, Jha et al.  1999).  Distance estimates have ranged from
9.6$\pm$0.6 Mpc from planetary nebulae (Feldmeier et al. 1997) to
11.6$\pm$0.9 Mpc from HST-Cepheid period luminosity estimates (Tanvir
et al. 1995). Subsequent HST-Cepheid period luminosity estimates place
M96 at 9.9 - 11.3 Mpc (Gibson \& Stetson 2001, 
Gibson et al.  2000, Hjorth \& Tanvir 1997), the 
range we will use in this study.  The CGRO
team was able to begin observations at about maximum light. A total of
88 days of observing by both the COMPTEL and the OSSE instruments were
devoted to SN 1998bu (spanning 17-136 days after the explosion), again
resulting in two separate data-sets. Neither instrument detected 847
or 1238 keV line emission. The OSSE instrument reported a 3$\sigma$ upper limit 
for 3~x~10$^{-5}$ phot~cm$^{-2}$~s$^{-1}$ for the 847 keV line based 
upon a combined fit to the 847 keV and 1238 keV lines (Leising at el. 1999). 
When treated separately, the derived formal fluxes are 
(1.2$\pm$1.4)~x~10$^{-5}$ phot~cm$^{-2}$~s$^{-1}$  and 
(-0.6~$\pm$1.6)~x~10$^{-5}$ phot~cm$^{-2}$~s$^{-1}$ for the 847 keV and 
1238 keV lines (M.D. Leising, personal communication). 
The COMPTEL 2$\sigma$ upper limits are
3.1~x~10$^{-5}$ phot~cm$^{-2}$~s$^{-1}$ and 2.3~x~10$^{-5}$
phot~cm$^{-2}$~s$^{-1}$ respectively (using the lower of the imaging
and spectral analysis results for each line, from Georgii et al. 2002).

\begin{figure}
\epsscale{0.7}
\centerline{\plotone{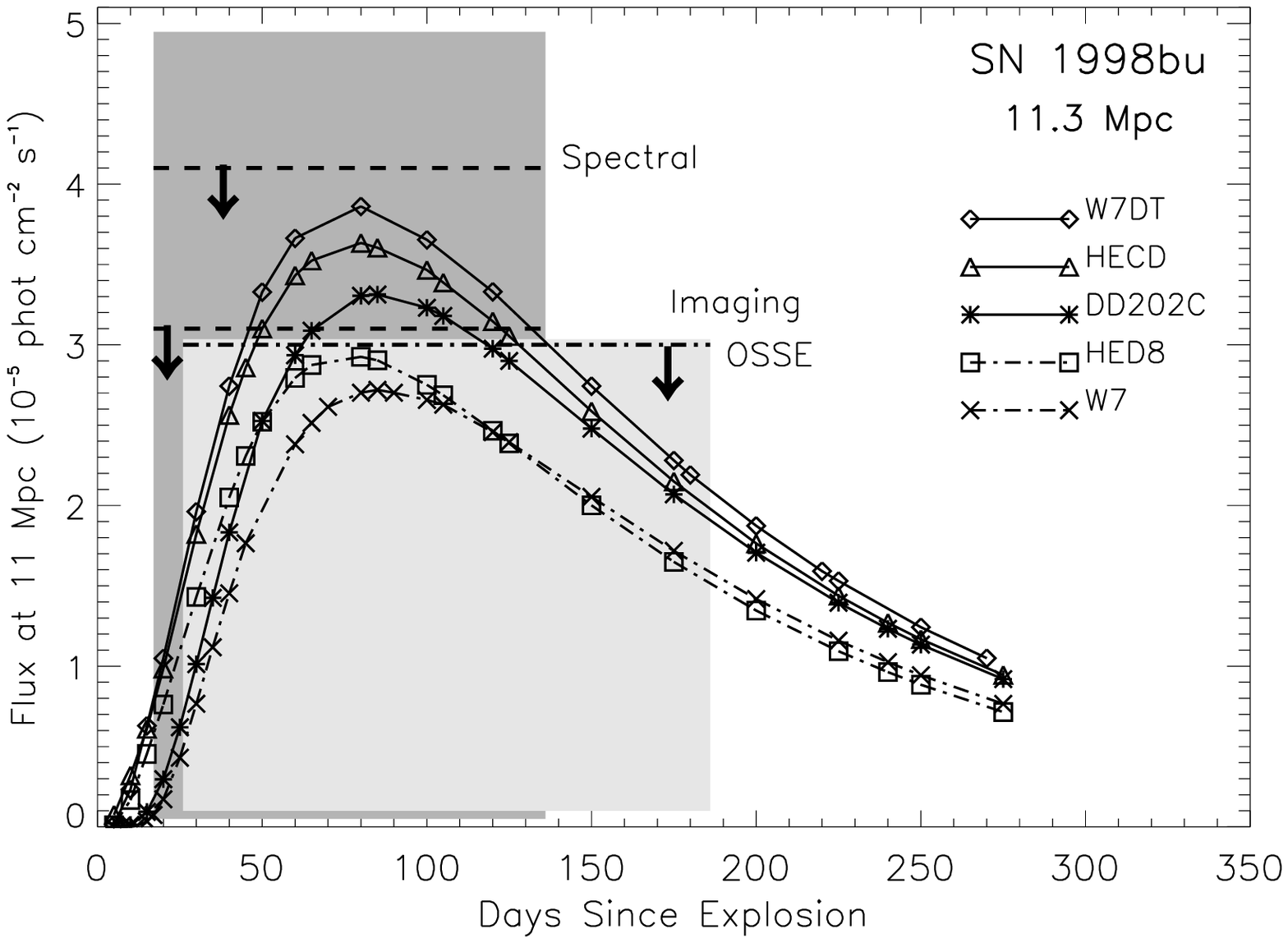}}
\centerline{\plotone{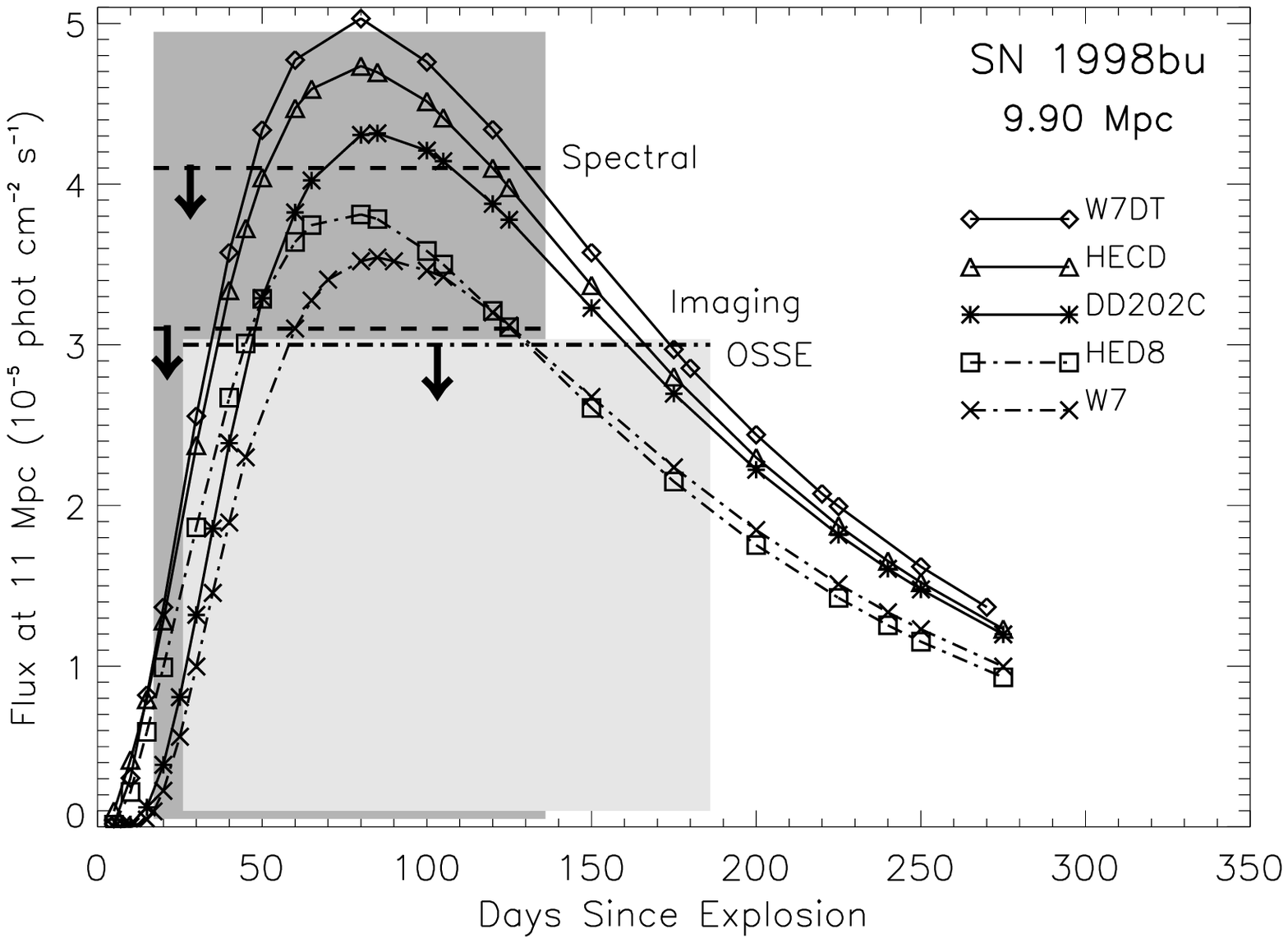}}
\caption{Five current simulations of 847 keV line emission from
SN Ia models at two assumed distances are compared with COMPTEL and OSSE 
observations of SN 1998bu. The upper panel shows the models at the larger distance, 
11.3 Mpc, the lower panel shows the models at 9.9 Mpc.  
The five current simulations are identified as shown.
The light shaded region shows roughly the epoch of OSSE observations, the dark shaded 
region shows roughly the epoch of COMPTEL observations. The two COMPTEL upper limits 
(Imaging and Spectral, dashed lines) are at the 2$\sigma$ level (Georgii et al. 2001). 
The OSSE 3$\sigma$ upper limits (dot-dashed line) are based upon a joint 847/1238 keV 
line fit. Table \ref{tab_obs} shows the
probabilities of each model being consistent with the data at the two distances.}     
\label{sn1998bu} 
\end{figure}

We first study the larger distance to NGC 3368.  Comparison of the
three normally-luminous SN models with these upper limits finds that
W7 and HED8 peak at or below the COMPTEL imaging upper limit for the
847~keV line, and that average flux of DD202C is approximately equal
to the COMPTEL imaging upper limit (Figure \ref{sn1998bu}). All three 
models are consistent with the combined OSSE and COMPTEL, 847 keV and 
1238 keV data at a 10\% probability, or better using the chi-squared 
test to the individual data points (Table \ref{tab_obs}). The
super-luminous SN models (HECD and W7DT) are brighter than the 
normally-luminous models, and are less likely to be 
as faint as the combined measurements. Considering that the
optical observations favor a normally-luminous SN Ia, the
non-detection is consistent with expectations.  

Assuming the shorter distance to NGC 3368, the upper limits become a
great deal more constraining. Only W7 appears to be faint enough
to rise above the 2\% probability level for having neither instrument 
detect emission from the SN. The gamma-ray observations appear to 
favor a larger distance to NGC 3368. 

Comparing these interpretations with Georgii et al. 2002, the
conclusions are similar, but not identical. Principally, at the larger
distance, that work only rejects the high-$^{56}$Ni producing models, while
at the shorter distance that work rejects all normally-luminous models. They
use the light curves shown in Kumagai \& Nomoto 1997, for HECD, W7DT,
W7 and WDD2, which were high as discussed in $\S4$.  The delayed
detonation light curves from that work 
are very similar to the DD202C light curve shown in
Figure \ref{sn1998bu}. We note that Table 1 in that work 
shows average fluxes that 
correspond to their shorter distance of 9.6 Mpc, not the 11.3 Mpc
shown in their Figure 6, and should thus be compared with our column 1
in Table \ref{tab_obs}.  It is also worth noting that the CGRO
observations spanned the epoch at which normally-luminous models
predict the brightest 847 and 1238 keV line emission. Thus, the
non-detection is not likely to have been affected by delay in the CGRO
observations.

\section{Conclusions}

In this paper, we compare gamma-ray emission simulations from 7
transport codes using a diverse set of SNe Ia models.  The spectra for
3 models (DD202C, W7, HED6) at explosion times ranging from 5 to over
200 days provide tests of these codes for a range of extreme
conditions.  This information allowed us to track down a number of
errors in past results and correct for these errors.  The results of
HWK98 and Kumagai \& Nomoto (1997) had the most dramatic errors, but their
revised ``current'' codes now agree much better with our ``unified''
solution.  

To the extent that 1D SN Ia models closely approximate the physical SN
explosion, observations can now be confidently compared with
simulations.  With current explosion scenarios and precise flux
measurements, sub-Chandrasekhar mass models can be clearly
distinguishable from Chandrasekhar mass models for normal and
sub-luminous SNe Ia.  However, with a suitably sensitive instrument, 
comparisons between line shapes, in addition to line fluxes, 
provide the best means to
distinguish different explosion scenarios (HWK98).  

Contrary to some of the past results, comparing to current data on SNe Ia 
finds that, for the sub-class of each explosion, theoretical gamma-ray
line fluxes from 1D models are consistent with the observations.
However, bear in mind that the explosion scenarios shown are limited by 
the adequacy of 1-dimensional modeling, and truly accurate comparisons 
will require 3-dimensional explosions and transport calculations.
In particular, clumping and global asymmetries will produce line profiles 
that differ from the profiles shown in this study. The wide range of 
line profiles possible with 3-dimensional simulations, and the resulting 
potential for confusion was partial motivation for this study. 

Finally, recall that the inverse of calculating the gamma-rays that
escape the SN ejecta (producing the gamma-ray flux) is the energy that
is deposited into the SN ejecta. The ability to simulate the
optical/IR/UV light curves of SNe Ia depends upon this deposition being
accurately treated. This project does not directly address the energy
deposition aspect of these simulations (and thus makes no claims), 
but errors in the decay rates and escape fraction 
may also lead to discrepancies in the energy deposition.  Gamma-ray transport, 
which provides the initial input for the emission of optical light, must 
be understood to model the optical light curves of supernovae.

The authors would like to thank Peter H\"{o}flich for assistance 
in running his gamma-ray transport code, MC-GAMMA, and Mark Leising 
for providing OSSE results from SNe 1991T and 1998bu. 

This work was performed under the auspices of the US Department of Energy
by Los Alamos National Laboratory, under contract W-7405-ENG-36, and by 
the National Science Foundation (CAREER grant AST 95-01634),
with support from DOE SciDAC grant number DE-FC02-01ER41176. 
L.-S. The acknowledges support from the DoE HENP Scientific 
Discovery through Advanced Computing Program."

\begin{table}
\caption{Characteristics of Seven Gamma-ray Transport Algorithms}
\begin{tabular}{|lc|cccccc|}
\hline
\hline

%Simulation & & Monte & Time & Line & Time & Positronium & Tag or  & Interactions \\
%Creator    & & Carlo & Dilation$^{a}$ & Broad$^{b}$ & Evolve$^{c}$ & 
%Fraction & Spectrum$^{d}$ & Treated$^{e}$\\
%or Name   &Ref. &$[y/n]$&$[y/n]$ & $[y/n]$ & $[y/n]$& f(Ps) & $[T/S]$ & $[CS,PE,PP]$ \\

Simulation & & Monte & Tag or & Bin Width      & Line        & Density & Positr. \\
Creator    & & Carlo & Spec.$^{a}$ & @ 847 keV & Broad$^{b}$ & Evolve$^{c}$ & Fraction \\ 
or Name   &Ref. &$[y/n]$& $[T/S]$ & [keV]   &$[y/n]$      & $[y/n]$ &  f(Ps) \\
           & &       &            &    & ($\S$ 2.1.1)& ($\S$ 2.1.2) & ($\S$ 2.2.2) \\ 

\hline
Boggs       &(1)& N & T & 2.8 & Y & Y & --- \\ 
Pinto       &(2)& Y & S & 2.4 & Y & Y & 0.0  \\
H\"{o}flich &(3)& Y & S & 2.4 & Y & N & 1.0  \\
Isern       &(4)& Y & S & 2.1 & Y & Y & 1.0  \\
Kumagai     &(5)& Y & T & 50  & N & N & 0.0  \\
Hungerford    &(6)& Y & T & 0.5 & Y & N & 0.0 \\ 
The         &(7)& Y & T & 40  & N & N & 1.0 \\ 
\hline
\end{tabular}
\begin{tabular}{|l|ccc|}
Simulation & Time             & Source        &Interactions\\
Creator &  Dilation$^{d}$ & Evolve$^{e}$ & Treated$^{f}$\\
        & $[y/n]$          &$[y/n]$        &$[CS,PE,PP]$\\
        & ($\S$ 2.2.3)     & ($\S$ 2.2.3)  & ($\S$ 2.3) \\
\hline
Boggs       &Y & Y & CS,PE \\
Pinto       & N & N & CS,PE,PP \\
H\"{o}flich & Y & N & CS,PE,PP \\
Isern & Y & N & CS,PE,PP \\
Kumagai & N & N & CS,PE,PP \\
Hungerford & N & N & CS,PE,PP \\
The & N & N & CS,PE,PP \\
\hline
\end{tabular}

\label{tab_sim}

\begin{tabular}{l}
$^{a}$ Is the line flux derived from determining the escape fraction of ``tagged" line photons, or \\
extracted from the spectrum and subject to line blending and continuum contamination? \\
$^{b}$ Are the photons emitted with Doppler broadening due to the differential expansion of the ejecta? \\
$^{c}$ Does the algorithm evolve the ejecta density after the photon emission to 
account for \\
non-zero crossing times? \\
$^{d}$ Are the relativistic effects of time dilation on the decay rate included? \\
$^{e}$ Does the algorithm account for the effect of requiring simultaneous photon arrival from \\
the near/far side of the ejecta? \\
$^{f}$ The interactions treated are CS = Compton Scattering, PE = Photoelectric Absorption, \\

PP = Pair Production. \\

REFERENCES. (1) Boggs (ApJ submitted) ; (2) Pinto, Eastman \& Rogers 2001; \\
(3) H\"{o}flich, Khokhlov \& Wheeler 1998; (4) Isern, Gomez-Gomar, \\
Bravo \& Jean 1997; (5) Kumagai 1996; (6) Hungerford et al. 2003; \\
(7) The \& Burrows 1991. \\
\end{tabular}
\end{table}

\normalsize
\begin{table}
\caption{Important Gamma-ray Line for $^{56}$Ni and $^{56}$Co 
Decays. Lines studied in this work are listed in bold font. 
All ratios are from the 8th Table of Isotopes (1996).}
\begin{tabular}{|lc|lc|}
\hline
\hline
\multicolumn{2}{|c|}{$^{56}$Ni Decay} & \multicolumn{2}{c|}{$^{56}$Co Decay} \\
\hline
Energy & Intensity & Energy & Intensity \\
$[keV]$ & $[phot/100 decays]$ & $[keV]$ &$ [phot/100 decays]$ \\
\hline
158 & 98.8 & {\bf 847} & {\bf 100} \\
270 & 36.5 & 1038 & 14 \\
480 & 36.5 & {\bf 1238} & {\bf 67} \\
750 & 49.5 & 1772 & 15.5 \\
{\bf 812} & {\bf 86.0} & 2599 & 16.7 \\
1562 & 14.0 &  3240$^{a}$ & 12.5 \\
\hline
\end{tabular}
\label{tab_dec}

\begin{tabular}{l}
a This line is the sum of a three line complex. \\
\end{tabular}

\vspace{1cm}
\begin{tabular}{|llcc|}
\hline
\hline
Source &  Reference &   $\tau$(Ni56) & $\tau$(Co56) \\
of Half-lives & and Year     &    [d]      &   [d]   \\
\hline
Nuc. Data Sheets & Junde 1999  & 6.075 & 77.233 \\
Table of Isotopes(8th)& Firestone 1996   & 5.9 & 77.27 \\
Table of Rad. Isotopes  & Browne \& Firestone 1986 & 6.10 & 77.7 \\
Table of Isotopes(7th) & Browne \& Dairiki 1978  & 6.10 & 78.8 \\
Table of Isotopes(6th) & Lederer \& Shirley 1967 & 6.1 & 77 \\
\hline
\end{tabular}
\end{table}

\begin{table}
\caption{Characteristics of SN Ia Explosion Models}
\begin{tabular}{|lccccc|}
\hline
\hline
Model & Mode of &  & M$_{*}$ & M$_{Ni}$ & E$_{kin}$ \\
Name & Explosion & Ref. & (M$_{\odot}$) & (M$_{\odot}$) & (10$^{51}$ ergs s$^{-1}$) \\
\hline
\multicolumn{6}{|l|}{Algorithm Comparison} \\
DD202C & Delayed det. & (1) & 1.40 & 0.72 & 1.33 \\
HED6 & He-det. &  (2) & 0.77 & 0.26 & 0.74 \\
W7 & Deflagration & (3)   & 1.37 & 0.58 & 1.24 \\
&   &       &     &      &     \\
\multicolumn{6}{|l|}{Spanning Explosions} \\
PDD54 & Pul.del.det. & (4) & 1.40 & 0.14 & 0.35 \\
W7DT  & Late det. & (5) & 1.37 & 0.76 & 1.61 \\
HED8 & He-det. & (2) & 0.96 & 0.51 & 1.00 \\
HECD & He-det. & (6) & 1.07 & 0.72 & 1.35 \\
&   &       &     &      &     \\
\hline
\end{tabular}
\label{tab_mods}

\begin{tabular}{l}
REFERENCES. (1) H\"{o}flich et al. 1998; (2) H\"{o}flich \& Khokhlov 1996;  \\
(3) Nomoto et al. 1984; (4) H\"{o}flich, Khokhlov \& Wheeler 1995; (5) Yamaoka \\ 
et al. 1992; (6) Kumagai 1997. \\
\end{tabular}
\end{table}

\begin{table}
\caption{Comparisons of SN Ia models with SN 1998bu}
\begin{tabular}{|lcccccc|}
\hline
\hline
%\multicolumn{7}{|l|}{{\bf SN 1991T$^{a}$}} \\
%&\multicolumn{3}{c}{Consistency at 11.3 Mpc} & \multicolumn{3}{c|}{Consistency at 14.0 Mpc} \\
%Model & OSSE & OCL & OCM & OSSE & OCL & OCM \\
%Name & [\%]  & [\%]  & [\%]  & [\%]  & [\%]  & [\%] \\
%\hline
%W7DT & 55.7  & 3.80  & 60.2  & 90.6  & 60.9  & 71.2  \\
%HECD & 63.4  & 7.6  & 63.8  & 92.6  & 69.9  & 70.9  \\
%W7   & 86.4  & 47.2  & 71.0  & 97.1  & 93.1  & 66.9  \\
%HED8 & 84.1  & 37.5  & 70.5  & 96.8  & 90.5  & 67.8  \\
%DD202C& 72.0 & 16.3  & 67.2  & 94.4  & 79.9  & 70.1  \\
%&&&&&& \\
%\hline
\multicolumn{7}{|l|}{{\bf SN 1998bu$^{a}$}} \\
Model & \multicolumn{2}{c}{$F_{10 Mpc}^{75 days}$} & \multicolumn{2}{c}{Consistency at 9.9 Mpc} & 
\multicolumn{2}{c|}{Consistency at 11.3 Mpc} \\
Name & \multicolumn{2}{c}{[$10^{-5} cm^{-2} s^{-1}$]} & \multicolumn{2}{c}{[\%]} & 
\multicolumn{2}{c|}{[\%]} \\
\hline
W7     & \multicolumn{2}{c}{3.3}  & \multicolumn{2}{c}{3.87}  & \multicolumn{2}{c|}{29.4} \\
HED8   & \multicolumn{2}{c}{3.7}  & \multicolumn{2}{c}{1.48}  & \multicolumn{2}{c|}{19.8} \\
DD202C & \multicolumn{2}{c}{4.1}  & \multicolumn{2}{c}{0.40}  & \multicolumn{2}{c|}{11.5} \\
W7DT   & \multicolumn{2}{c}{4.6}  & \multicolumn{2}{c}{0.05}  & \multicolumn{2}{c|}{4.73} \\
HECD   & \multicolumn{2}{c}{4.9}  & \multicolumn{2}{c}{0.02}  & \multicolumn{2}{c|}{2.84} \\
\hline
\end{tabular}
\label{tab_obs}
\begin{tabular}{l}
%$^{a}$ OCL = OSSE data plus Lichti et al. 1994 analysis of COMPTEL observations. \\
%OCM = OSSE data plus Morris et al. 1998 analysis of COMPTEL observations. \\
%OSSE data from Leising et al. 1994. \\
$^{a}$ OSSE data from Leising et al. 1999, COMPTEL data from Georgii et al. 2002. \\
\end{tabular}
\end{table}

%\begin{figure}
%\epsscale{1}
%\plotone{ps/latespec100.ps}
%\caption{Line profiles of the 847 keV line for the SN Ia model, DET, 
%at 100 days.  The line appears flattened in both simulations. }
%\label{sawtooth}
%\end{figure}

\end{document}